\documentclass[pra,showpacs,twocolumn]{revtex4-1}
\usepackage{bm,graphicx,amsmath}

\newcommand{\beq}{\begin{equation}}
\newcommand{\enq}{\end{equation}}
\newcommand{\la}{\langle}
\newcommand{\ra}{\rangle}

\begin{document}
\title{Multi-orbital bosons in bipartite optical lattices}
\author{Jani-Petri Martikainen$^{1,2}$}
\author{Jonas Larson$^{3,4}$}
\affiliation{$^1$COMP Centre of Excellence, Department of Applied Physics, Aalto University, Fi-00076 Aalto, Finland}
\affiliation{$^2$NORDITA, Se-106 91 Stockholm, Sweden}
\affiliation{$^3$Department of Physics, Stockholm University, 
Se-106 91 Stockholm, Sweden}
\affiliation{$^4$Institut f\"ur Theoretische Physik, Universit\"at zu K\"oln, K\"oln, De-50937, Germany}
\date{\today}

\begin{abstract}
We study interacting bosons in a two dimensional square bipartite 
optical lattice.
By focusing on the regime where the first three excited bands
are nearly degenerate we derive a three orbital tight-binding model 
which captures the most relevant features of the bandstructure
when the first excited $p$-bands in another sublattice are nearly degenerate
with $s$-band of the other sublattice.
In addition, we also derive a corresponding generalized Bose-Hubbard model and
solve it numerically under different situations, both with and without a confining trap. It is especially found that the hybridization between sublattices can strongly influence the phase diagrams and in a trap enable even appearances of condensed phases intersecting the same Mott insulating plateaus.
\end{abstract}


\maketitle

\section{Introduction}
The understanding that Hubbard models can be realized with ultracold atoms
in optical lattices~\cite{Jaksch1998a} has stimulated extensive
effort to explore different aspects of quantum many-body physics
in optical lattices~\cite{Bloch2008a,Lewenstein2007a}.
The early works focused on the lowest energy band and in a pioneering
experiment by Greiner {\it et al.}~\cite{Greiner2002a} the Mott-superfluid transition with ultracold bosons was observed. More recently, experimental groups have started to probe the properties of ultracold atoms under circumstances where the excited energy bands~\cite{Lewenstein2011a} can no longer be ignored. This is most relevant since it has been demonstrated that the emerging multi-orbital effects can indeed have crucial effects also on the ground state phase diagrams~\cite{Soltan-Panahi2012a}. These excited bands can become important either when the atom-atom interactions become very large~\cite{Kohl2006a,Larson2009a,Will2010a,Buchler2010a,Hazzard2010a,Mering2011a,Von_Stecher2011a,Bissbort2011a}, or when atoms are
deliberately prepared on the excited bands. Such 'out of equilibrium' state preparation has been established by using accelerating lattices~\cite{Browaeys2005a} or Raman transitions between bands~\cite{Mueller2007a}. In the realm of these new experiments, one hopes to explore the regime where meta stable excited many-body states show very different properties from those of the ground state~\cite{Isacsson2005a,Scarola2005a,Scarola2006a,CongjunWu2006a,Liu2006a,Xu2007a,Collin2010a}.

The experiment most closely relevant for our purposes is the one
by Wirth {\it et al.}~\cite{Wirth2011a}. Bosonic atoms were prepared in the ground state of a bipartite optical lattice and then the lattice was suddenly changed so that 
the initial ground state band atoms became (quasi) degenerate with 
a set of other bands which were initially separated by a large
band gap. This process drove atoms into bands with non-trivial 
orbital properties and enabled the observation of superfluidity on these so called $p$-bands. This experiment was followed by others~\cite{Olschlager2011a,Olschlager2011b} where unconventional superfluidity was observed in the even more excited $f$-bands.

Motivated by these experiments and especially on 
the aspects of the physics when different bands become degenerate
we study multi-band bosons in a bipartite square lattice when
bands cross.
Such band crossing can imply topologically non-trivial bandstructures~\cite{CastroNeto2009a,Qi2011a}.
In principle, with the help of artificial gauge fields such bandstructures 
can also be engineered on the lowest band~\cite{Lim2008a,Larson2010a}, but they 
might be easier to engineer in the excited bands were artificial 
gauge fields may become unnecessary. For example, in a square bipartite 
lattice the bandstructure can be composed of flat bands intersecting Dirac cones which, on the one hand, have interesting analogs with graphene physics, but 
the flat bands also have novel influences on the dynamical properties of the gas~\cite{Apaja2010a,Hyrkas2012a}. 
Physics of Dirac fermions have been studied in square optical lattices
also in the absence of the flat band~\cite{Kennett2011a}. 

As in the experiment by Wirth {\it et al.}~\cite{Wirth2011a}, we consider a bipartite square lattice of deep $\mathcal{A}$-sites and more shallow $\mathcal{B}$-sites which however have a higher energy offset. Under such circumstances, the excited (localized) states in $\mathcal{A}$ sites can become resonant with the ground states in $\mathcal{B}$ sites. When this happens, the $p$-bands can be strongly
hybridized with the $d$-band. For vanishing atom-atom interaction, most of the relevant physics is captured by a tight-binding (TB) model, which predicts the existence
of Dirac points and a flat band. Proceeding by adding atom-atom interactions we
derive a generalized multi-band Bose-Hubbard model. We solve this theory from weak to strong interactions as well as in a trap. 
The calculated solutions reveal transitions from incompressible Mott insulators to condensed phases, but due to different atom-atom interactions the Mott lobes can be very dissimilar from those predicted by the usual single-band Bose-Hubbard model. Furthermore, the solution in a trap reveals the possibility that condensed states
in different sublattices occur in different regions of the trap. 
Our findings complement some other very recent ones, like Ref.~\cite{Shchesnovich2012a} where $p$-band bosons in a shallow bipartite optical lattice in terms of a nonlinear boson model is studied, and the work~\cite{Cai2011a} analyzing the band structure renormalized by the presence of interactions and the condensate in the broken symmetry phase. Finally, Sun {\it et al.}~\cite{Sun2011a} also derived a fermionic tight-binding model which is 
quite similar to the one used by us.

The paper is organized as follows. We begin by outlining the 
theory relevant for our purposes in Sec.~\ref{sec:theory}. In particular, Sec.~\ref{sec:idealtheory} presents the tight-binding model to describe the ideal gas of atoms
and in Sec.~\ref{sec:interactingtheory} we extend the model to include atom-atom
interactions. In Sec.~\ref{sec:gutzwillerresults} the generalized Bose-Hubbard model is solved within the Gutzwiller ansatz approach, and in Sec.~\ref{sec:trappedsystem} we discuss the solution in a harmonic trap. We end with a few concluding remarks in Sec.~\ref{sec:conclusions}.

\section{Theoretical formulation}
\label{sec:theory}
\subsection{Ideal system}
\label{sec:idealtheory}
\begin{figure}
\includegraphics[width=0.5\textwidth]{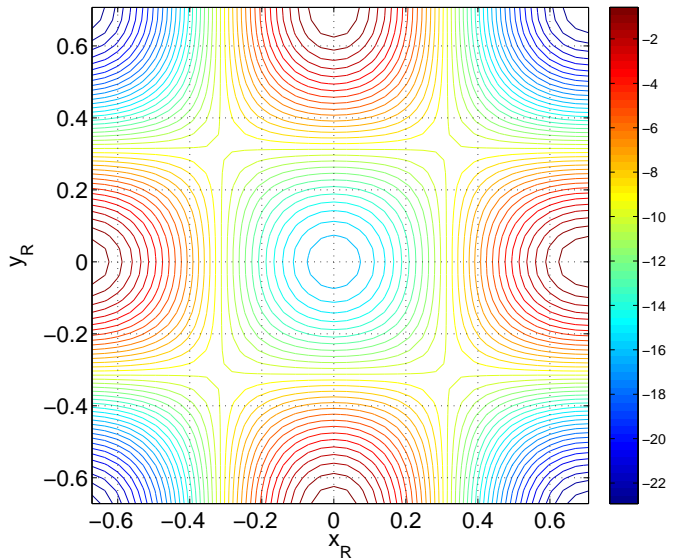}
\caption{(Color online) The symmetric lattice potential with $V_0=10\,E_R$ 
over one unit cell. The parameters were chosen as $\epsilon=\eta=1$,
$\alpha=0$, and $\theta/\pi=0.556$. $x_R$ and $y_R$ refer to coordinate
axis rotated by $\pi/4$ with respect to the laboratory axes $\hat{x}$ and $\hat{y}$. The shallow $\mathcal{B}$-site is in the center while
the deeper $\mathcal{A}$-sites are in the corners. Distance, $\lambda/2$, between $\mathcal{A}$- and $\mathcal{B}$-sites was taken as a unit of length.
}
\label{fig:potential}
\end{figure}

\begin{figure*}
\includegraphics[width=0.35\textwidth]{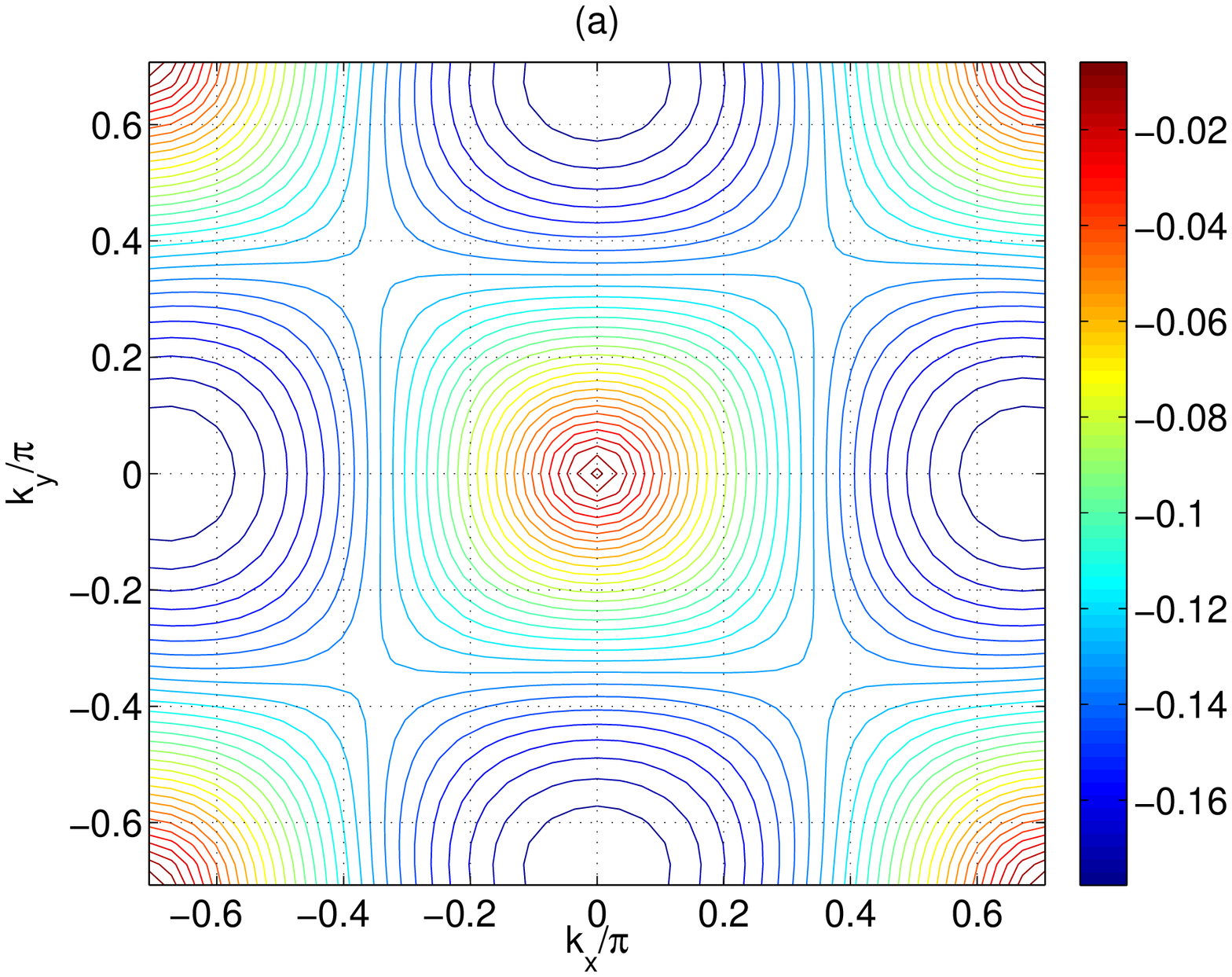}
\includegraphics[width=0.33\textwidth]{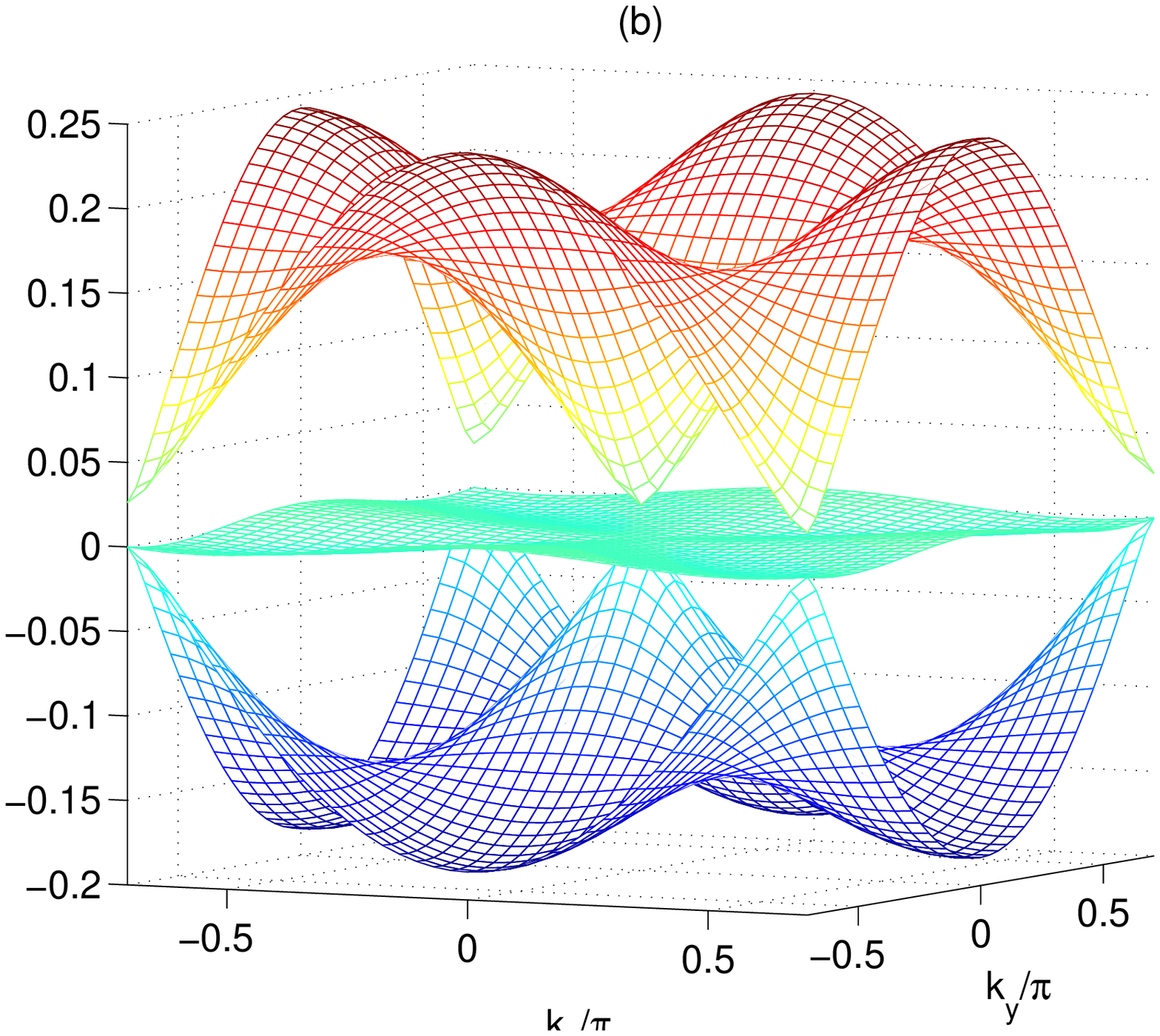}\\
\includegraphics[width=0.35\textwidth]{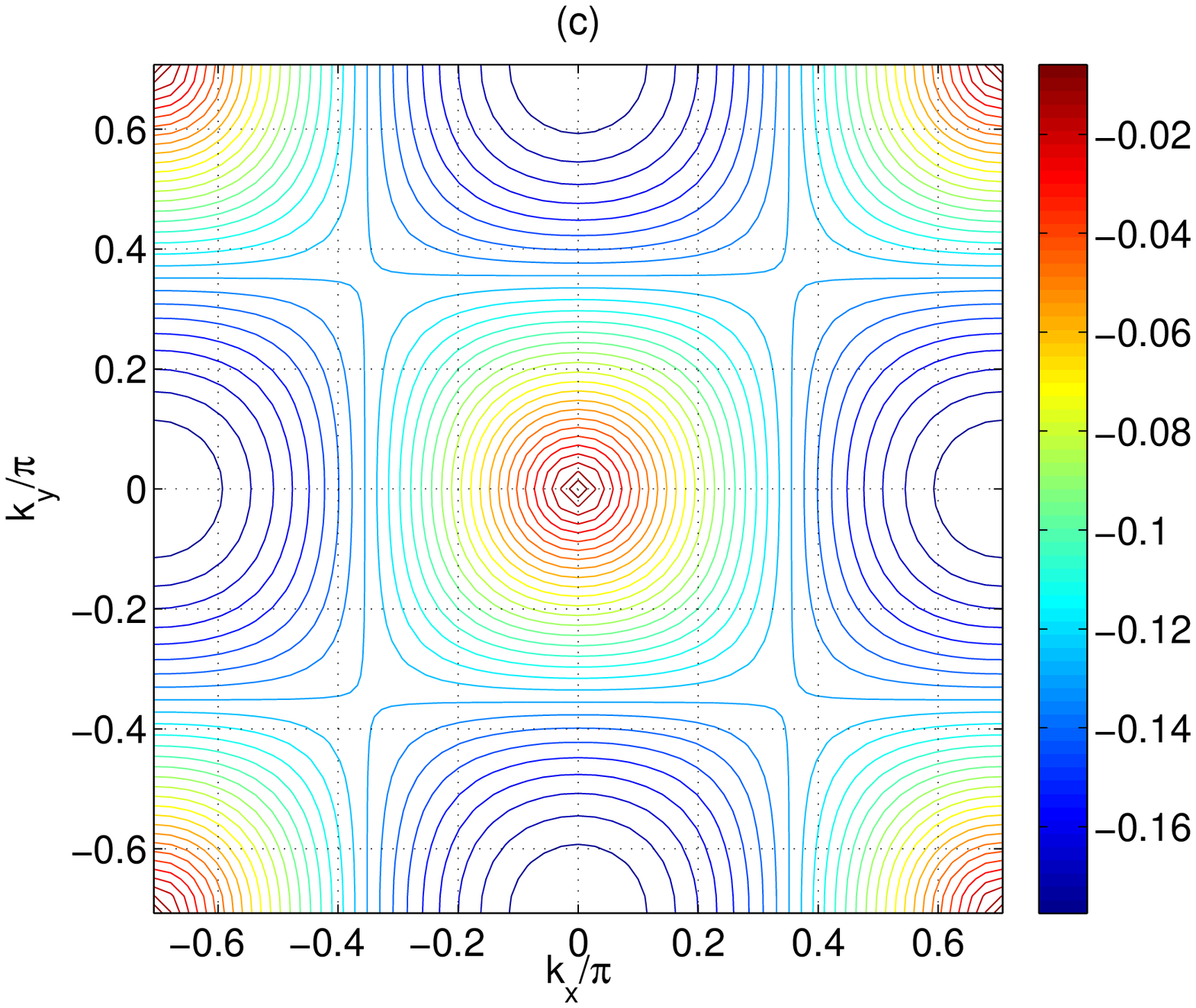}
\includegraphics[width=0.33\textwidth]{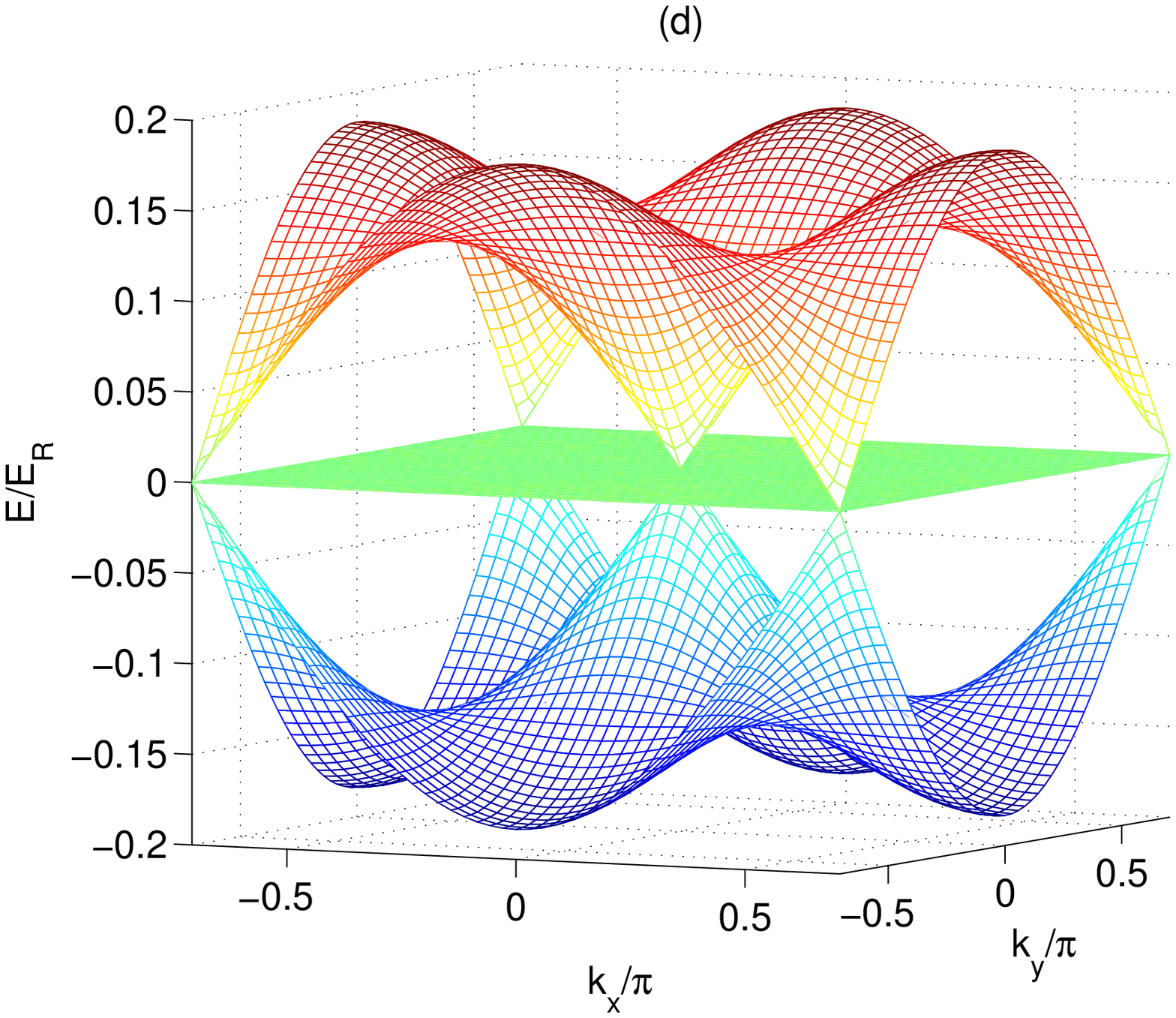}
\caption{(Color online) Dispersions of the three lowest excited bands.
(a) and (b) are numerically calculated for a lattice with $V_0=10\,E_R$,
$\epsilon=\eta=1$, $\alpha=0$, and $\theta/\pi=0.556$. (a) shows the lowest
excited band while (b) shows all three excited bands in the same plot.
(c) and (d) are calculated with the TB model with parameters
$t_{xx}^{AB}=t_{yy}^{AB}=0.06485E_R$ and $\delta=0$.
}
\label{fig:dispersiondemo}
\end{figure*}

We will assume a two-dimensional lattice potential similar 
to the one used in the experiments by Wirth {\it et al.}~\cite{Wirth2011a};
\beq
\label{eq:potential}
\begin{array}{lll}
V(x,y)\! & = & \displaystyle{\!- \frac{V_0}{4} \Big|\eta\left[  \left(\hat z \,\cos{(\alpha)}  + \hat y \sin{(\alpha)} \right)e^{i k x}  + \epsilon \hat z e^{-i k x} \right]} \\ \\
& & + \, e^{i \theta} \hat z \left(e^{i k y} + \epsilon \, e^{-i k y} \right) \Big|^2 \,,
\end{array}
\enq
where $V_0$ is the lattice depth, $k$ the lattice wave number, $\eta$ accounts for a small difference in 
the powers directed to different interferometer branches, 
$\epsilon$ characterizes the power reduction 
in the retro-reflected beams due to imperfect optics, 
and the angle $\alpha$ tunes the anisotropy introduced if $\epsilon \neq 1$. The angle $\theta$ sets a relative phase between the two standing waves. $\hat x, \hat y,$ and $\hat z$ are the unit vectors in the respective directions. Furthermore, the transverse $\hat z$-direction has been reduced due to tight confinement.  We will mostly consider a symmetric lattice with $\epsilon=\eta=1$,
and  $\cos{(\alpha)}= \epsilon$, but since different parameter choices can break the $p$-band degeneracies we allow for such possibilities as well. In Fig.~\ref{fig:potential} we show an example of a unit cell of this potential. Generally, the lattice is a bipartite square lattice where the two sublattices have lattice sites of different depths. Here we are interested in the parameter regime where the ground state in the shallow sites is quasi resonant with the first excited states of the deep sites. The resulting bandstructure of the regime we are interested in is depicted in Fig.~\ref{fig:dispersiondemo} (a) and (b). Here, and in the following, we scale the energies in terms of the recoil energy $E_R=\hbar^2(2\pi/\lambda)^2/2m$ of the atoms with mass $m$ to absorb a photon of wavelength $\lambda$. In particular, Fig.~\ref{fig:dispersiondemo} is calculated for $V_0=10E_R$. In this region, the two lowest excited p-bands become degenerate with the $d$-band. When this happens non-trivial bandstructures with Dirac points emerge. Furthermore, one of the bands is almost flat suggesting that interactions play a larger role for atoms prepared in this band.

Restricting our analysis to the three bands of Fig.~\ref{fig:dispersiondemo}, i.e.
the localized ground state in the shallow $\mathcal{B}$-sites and the first two excited states in deep $\mathcal{A}$-sites, we obtain an effective theory in terms of three different orbitals.
In the absence of an external trap we can write the ideal gas Hamiltonian
in momentum space as 
\beq
H=\sum_{{\bf k}} \phi_{\bf k}^\dagger \hat{H}(k) \phi_{\bf k},
\enq
where 
\beq
 \phi_{\bf k}=\left[\begin{array}{c}
\hat\psi_{s,{\bf k}}^{\mathcal{B}}
\\
\hat\psi_{x,{\bf k}}^{\mathcal{A}}
\\
\hat\psi_{y,{\bf k}}^{\mathcal{A}}
\end{array}\right]
\enq 
describes the three types of orbitals included in our theory. There is
an $s$-like orbital in the shallow $\mathcal{B}$-sites, $\hat\psi_{s,{\bf k}}^{\mathcal{B}}$ and $p$-like $x$- and $y$-orbitals in the deep $\mathcal{A}$-sites, $\hat\psi_{x,{\bf k}}^{\mathcal{A}}$ and $\hat\psi_{y,{\bf k}}^{\mathcal{A}}$ respectively. When the energy of the $s$-orbital in the $\mathcal{B}$-sites is close to the energy of the $p$-orbitals in the $\mathcal{A}$-sites, the dominant tunneling process is the one hybridizing orbitals in different sublattices. This involves nearest neighbors and lower barrier height for tunneling while other tunneling processes require couplings over larger distances
and are therefore greatly suppressed. Thus, for sufficiently deep lattices
we can ignore tunnelings within $\mathcal{A}$- or $\mathcal{B}$-sites. On the other hand, since they only involve single particle physics, our theory can naturally include next nearest neighbor tunnelings easily when those are required.
 
In momentum space, this results in a TB model
\beq
\hat{H}(k)\!=\!\left[\!\begin{array}{ccc}
E_s^{\mathcal{B}}({\bf k}) & -2it_{xx}^{\mathcal{AB}}\sin\left(k_x\right) 
& -2it_{yy}^{\mathcal{AB}}\sin\left(k_y\right)  \\
2it_{xx}^{\mathcal{AB}}\sin\left(k_x\right)  & E_x^{\mathcal{A}}({\bf k}) & 0  \\
2it_{yy}^{\mathcal{AB}}\sin\left(k_y\right) & 0 & E_y^{\mathcal{A}}({\bf k}),
\end{array}\!\right]
\label{eq:TBAHamiltonian}
\enq
whose parameters can be deduced from the exact band structure calculations, see Fig.~\ref{fig:dispersiondemo}. In the next section, this model will also be given in position space. One consequence of the hybridization can be seen in how the orbital character of the system enters for example in the $\sin$-terms in the above TB model. Hopping occurs between $s$- and $p$-orbitals, which implies that the tunneling coefficient alternates signs between neighboring sites giving rise to a $\sin$- rather than a $\cos$-dispersion. In order to simplify notations, we choose our zero energy level to be the energy of the $s$-orbital in the $\mathcal{B}$-sites. 
Since only nearest neighbor tunneling processes are included,
the momentum dependence disappears from the 
diagonal terms and we have $E_s^{\mathcal{B}}({\bf k})=0$, $E_x^{\mathcal{A}}({\bf k})=E_x^{\mathcal{A}}\equiv\delta/2$, and $E_y^{\mathcal{A}}({\bf k})=E_y^{\mathcal{A}}\equiv-\delta/2$.

We note that a somewhat related TB model was also derived
by Sun {\it et al.}~\cite{Sun2011a}. However, in that model the
underlying lattice potential was different and the $p$-orbitals
were degenerate while in our case they can be different to account
for the possible anisotropy of $\mathcal{A}$-sites. This anisotropy was indeed an important ingredient in the experiment by Wirth {\it et al.}~\cite{Wirth2011a}.
In the symmetric case with $\delta=0$ the lowest energy state
of the TB model is $4$-fold degenerate, but this degeneracy is lifted
as soon as $\delta\neq 0$ so that the minima is only two-fold degenerate.

Furthermore, in the symmetric case with $\delta=0$ the sin-dispersions give rise to Dirac points at the origin as well as on the edges of the first Brillouin zone at $(\pm\pi/\sqrt{2},0)$ and $(0,\pm\pi/\sqrt{2})$. A non-zero detuning $\delta$ implies an effective mass term that split the Dirac point degeneracies. Similarly, in graphene the relativistic electrons become massive when the symmetry between the corresponding two triangular sublattices is broken~\cite{CastroNeto2009a}. Contrary to graphene, rather than having a two-level structure, the present model has three bands and the Berry phase as a Dirac point is encircled vanishes. 
The additional level appears as a flat band sandwiched between the other two bands. 

$\hat{H}(k)$ of  Eq.~(\ref{eq:TBAHamiltonian}) has the same structure as the Hamiltonian for a $\Lambda$-scheme frequently occurring in light-matter interaction models in quantum optics, and we can directly conclude that states of the flat band correspond to dark states with zero energy. These eigenstates are superpositions of $p$-orbitals and have a vanishing amplitude of being in the (``excited'') $s$-state in $\mathcal{B}$-sites~\cite{Arimondo1996a}. With this in mind, by considering anisotropic lattices ($t_{xx}^\mathcal{AB}\neq t_{yy}^\mathcal{AB}$) we notice that it would be possible to apply various examples of complete or fractional {\it stimulated Raman adiabatic passage} schemes~\cite{Bergmann1998a} 
to prepare specific orbital states for the atoms. Intriguingly, the Hamiltonian in Eq.~(\ref{eq:TBAHamiltonian}) also has a clear connection to spin-orbit coupled systems. In the long wavelength limit we can expand the trigonometric functions and find that the coupling between orbitals is linearly proportional to momentum~\cite{Qi2008a,Hasan2010a,Dalibard2011a} .
Usually spin-orbit coupling in ultracold atom systems is generated between different atomic hyperfine states~\cite{Lin2011a,Dalibard2011a}. Here the internal states of the atoms are not effected, but the spin-orbit-like coupling is a bandstructure effect that occurs between different orbitals.

In Fig.~\ref{fig:dispersiondemo} (c) and (d) we demonstrate that the TB model above is indeed a good approximation close to band degeneracy by comparing it with the numerically calculated bandstructure, plots (a) and (b). As can be seen, for the symmetric lattice
it reproduces the main features of the real bandstructure very well. Corrections beyond nearest neighbor hopping terms is seen to give rise to higher order variations in the dispersions mostly clear in the flat band. The tunneling coefficients $t_{xx}^{\mathcal{AB}}$ and $t_{yy}^\mathcal{AB}$ have been extracted from the band widths of the numerically obtained bands. While our model does works well close to resonance, it should be kept in mind that generally the real bandstructure is more complicated and more tunneling processes might have to be included in the theory.

\subsection{Interacting system}
\label{sec:interactingtheory}
In the previous section the ideal gas theory was derived and we now proceed by adding the atom-atom interactions. For ultracold atoms, interactions can be well modeled by contact interactions,
\beq
U=\frac{g}{2}\int d{\bf r}\hat\psi^\dagger({\bf r})
\hat\psi^\dagger({\bf r})\hat\psi({\bf r})\hat\psi({\bf r}).
\enq
In a deep lattice the field operator $\hat\psi({\bf r})$ is naturally
expanded in terms of the localized orbitals described by the Wannier
wave-functions $w_x^\mathcal{A}(x,y)$, $w_y^\mathcal{A}(x,y)$, and $w_s^\mathcal{B}(x,y)$. That is, we truncate the Hilbert space to contain only the three most relevant bands, i.e. the expansion is restricted to $\mathcal{B}$-sites' $s$-orbitals and $\mathcal{A}$-sites' $p$-orbitals. 

In the usual way, we limit the interaction to include only the dominant onsite terms. The strengths of various interactions are proportional to the scattering length, but their relative magnitudes depend on the orbital wavefunctions. To estimate these strengths we approximate the onsite orbitals with harmonic oscillator
wave-functions and in this way can analytically solve the integrals describing
interaction between $x$-orbitals in $\mathcal{A}$-sites
\beq
U_{xx}=U_0\int dxdy |w_x^\mathcal{A}(x,y)|^4,
\enq
between $y$-orbitals in $\mathcal{A}$-sites
\beq
U_{yy}=U_0\int dxdy |w_y^\mathcal{A}(x,y)|^4,
\enq
between $x$- and $y$-orbitals in $\mathcal{A}$-sites
\beq
U_{xy}=U_0\int dxdy |w_x^\mathcal{A}(x,y)|^2|w_y^\mathcal{A}(x,y)|^2,
\enq
and finally between $s$-orbitals in $\mathcal{B}$-sites
\beq
U_{sB}=U_0\int dxdy |w_s^\mathcal{B}(x,y)|^4.
\enq
We take that the remaining prefactor $U_0$ is tunable either 
by changing the lattice depth or by changing the effective scattering length.
In the harmonic approximation $U_{xy}=U_{xx}/3$. This condition 
can sometimes lead to accidental degeneracies, which are removed
as soon as the condition is broken~\cite{Collin2010a}. 
However, in this work this does not play a major role. Since 
the shallow sites are wider than the deep sites, their
orbitals are also more extended. This implies that $U_{sB}$ is often 
surprisingly close to the values of $U_{xx}$ and $U_{yy}$ even though these involve wider excited state orbitals. For concreteness, in the following we choose the lattice depth as $V_0=10\,E_R$ in which case it turns out that $U_{xx}=U_{yy}\approx 0.95\,U_{sB}$.

With the above introduced interaction strengths, we are now in a position to write down a many-body Hamiltonian describing multi-orbital bosons in a bipartite optical lattice. The corresponding Hamiltonian takes the form
\beq
H_{T}=H_0+H_{I,\mathcal{B}}+H_{I,\mathcal{A}},
\enq
where
\begin{equation}
\begin{array}{lll}
H_0&=&\displaystyle{\frac{\delta}{2}\sum_{{\bf i}\in \mathcal{A}} \left(\hat n_{x,{\bf i}}^\mathcal{A}-\hat n_{y,{\bf i}}^\mathcal{A}\right)}\\ \\
& & \displaystyle{-\frac{1}{2}\sum_{\alpha\beta}\sum_{\langle{\bf i},{\bf j}_{\beta_+}\rangle}
\left(t_{\alpha\beta}^{\mathcal{AB}}\hat\psi_{s,{\bf j}_{\beta_+}}^{\mathcal{B}\dagger}\hat\psi_{\alpha,{\bf i}}^{\mathcal{A}}
+h.c.\right)}\\ \\
& & \displaystyle{+\frac{1}{2}\sum_{\alpha\beta}\sum_{\langle{\bf i},{\bf j}_{\beta_-}\rangle}
\left(t_{\alpha\beta}^{\mathcal{AB}}\hat\psi_{s,{\bf j}_{\beta_-}}^{\mathcal{B}\dagger}\hat\psi_{\alpha,{\bf i}}^{\mathcal{A}}
+h.c.\right)}\\ \\
& &\displaystyle{-\mu \sum_{{\bf i}\in \mathcal{A}} \left(\hat n_{x,{\bf i}}^\mathcal{A}+\hat n_{y,{\bf i}}^A\right)
-\mu \sum_{{\bf i}\in \mathcal{B}}\hat n_{s,{\bf i}}^\mathcal{B}}
\end{array}
\label{eq:H0}
\end{equation}
describes the energy offsets and
nearest-neighbor tunneling giving rise to hybridization
between orbitals. Here ${\bf i}=(i_x,i_y)$ labels the lattice sites and $\mu$ is
the chemical potential. $\hat n_{x,{\bf i}}^\mathcal{A}$,
$\hat n_{y,{\bf i}}^\mathcal{A}$, and $\hat n_{s,{\bf i}}^\mathcal{B}$ are the number 
operators for $x$- and $y$-orbitals in an $\mathcal{A}$-site ${\bf i}$ and
$s$-orbitals in a $\mathcal{B}$-site ${\bf i}$.
The notation ${\bf j}_{\beta_+}$ (${\bf j}_{\beta_-}$)
indicates a nearest neighbor
of ${\bf i}=(i_x,i_y)$ 
to the right (left) in the direction $\beta\in \{\hat x,\hat y\}$.
For example, ${\bf j}_{x_+}=(i_x+1,i_y)$ while ${\bf j}_{x_-}=(i_x-1,i_y)$.
Finally, $h.c.$ indicates the hermitian conjugate. 
The hopping term must be written in this way since 
in this case tunneling is sensitive to the left and right difference.
Intuitively this is easy to understand by considering a $p$-orbital with a 
node. This orbital wave function changes sign as one moves along to axis
towards the neighboring site with $s$-orbital wavefunction.  
The overlap of these two
wavefunctions is predominantly positive if the neighbor is to the left
(for example), but predominantly negative if it is to the right. Note how such ``space-dependence'' in the hopping term also appears in lattice models exposed to (synthetic) magnetic fields~\cite{Dalibard2011a}.  
The tunneling parameters $t_{\alpha\beta}^{\mathcal{AB}}$ denotes the strength of tunneling of $\alpha$-orbitals in the $\mathcal{A}$ sublattice in the direction 
$\beta$ into the nearest neighbor $s$-orbital in the $\mathcal{B}$ sublattice. In the theory used here $t_{xy}^{\mathcal{AB}}=t_{yx}^{\mathcal{AB}}=0$. In the momentum representation, the term $H_0$ corresponds to the TB Hamiltonian encountered in the previous subsection.

The remaining terms describe interactions.
\beq
H_{I,\mathcal{B}}=\frac{U_{s\mathcal{B}}}{2}\sum_{{\bf i}\in \mathcal{B}} 
\hat n_{s,{\bf i}}^\mathcal{B}\left(\hat n_{s,{\bf i}}^\mathcal{B}-1\right)
\label{eq:HB}
\enq
accounts for the interactions in the $\mathcal{B}$-sites and
\begin{equation}
\begin{array}{lll}
H_{I,\mathcal{A}}&=&\displaystyle{\sum_{{\bf i}\in \mathcal{A}}\left[ \frac{U_{xx}}{2}\hat n_{y,{\bf i}}^\mathcal{A}\left(\hat n_{y,{\bf i}}^\mathcal{A}-1\right)
+\frac{U_{yy}}{2} \hat n_{y,{\bf i}}^\mathcal{A}\left(\hat n_{y,{\bf i}}^\mathcal{A}-1\right)\right]} \\ \\
& & +\displaystyle{\frac{U_{xy}}{2}\left[\hat\psi_{x,{\bf i}}^{\mathcal{A}\dagger}\hat\psi_{x,{\bf i}}^{\mathcal{A}\dagger}
\hat\psi_{y,{\bf i}}^{\mathcal{A}}\hat\psi_{y,{\bf i}}^{\mathcal{A}}+
\hat\psi_{y,{\bf i}}^{\mathcal{A}\dagger}\hat\psi_{y,{\bf i}}^{\mathcal{A}\dagger}
\hat\psi_{x,{\bf i}}^{\mathcal{A}}\hat\psi_{x,{\bf i}}^{\mathcal{A}}\right]}\\ \\
& & +2U_{xy} \hat n_{x,{\bf i}}^\mathcal{A}\hat n_{y,{\bf i}}^\mathcal{A} 
\label{eq:HA}
\end{array}
\end{equation}
for the interactions within the $\mathcal{A}$-sites.
This term is somewhat more complicated than the corresponding term in the shallow
sites since $x$- and $y$-orbitals interact and (for bosons)
can change into one another. 
Finally, we note that when  $\delta=0$
the total Hamiltonian supports a symmetry corresponding to swapping of $x$- and $y$-flavor atoms in the $\mathcal{A}$-sites.

\section{Gutzwiller results}
\label{sec:gutzwillerresults}
The Gutzwiller ansatz~\cite{Zakrzewski2005a}  
for the many-body wave functions provides a reasonably
accurate description of interacting bosonic systems, especially in dimensions $D>1$. 
Due to the bipartite lattice and multiple flavors in one sublattice, our case
is somewhat more complex than the usual Bose-Hubbard model. 
The Gutzwiller ansatz we use is given by
\beq
|\psi\ra=\prod_{\bf i\in \mathcal{A}}\sum_{{\bf n}^\mathcal{A}}
a_{{\bf n}^\mathcal{A}}^{({\bf i})} |{\bf n}^\mathcal{A}\ra_{\bf i}
\prod_{\bf j\in \mathcal{B}}\sum_{n_s^\mathcal{B}}
b_{n_s^\mathcal{B}}^{({\bf j})} |n_s^\mathcal{B}\ra_{\bf j}.
\label{eq:Guztwiller}
\enq
The expansion 
coefficients $a_{{\bf n}^\mathcal{A}}^{({\bf i})}=a_{n_x^\mathcal{A},n_y^\mathcal{A}}^{({\bf i})}$ 
and $b_{n_s^\mathcal{B}}^{({\bf j})}$ 
are the 
Gutzwiller amplitudes of the corresponding on-site Fock state. 
For our purposes, in the $\mathcal{A}$-sites the relevant subspace is spanned by the Fock states of the form $|{\bf n}^\mathcal{A}\ra=|n_x^\mathcal{A},n_y^\mathcal{A}\ra$, where  $n_\alpha^\mathcal{A}$ is the occupation
number of the $\alpha$-orbital. In the $\mathcal{B}$-sites the wave function is
expanded in terms of Fock states $|n_s^\mathcal{B}\ra$ associated with 
$s$-orbitals. The Gutzwiller ansatz captures the onsite physics exactly,
but ignores some correlations between sites. In the limit where
the onsite wave function are taken to be coherent states it recovers
the Gross-Pitaevskii limit of weakly interacting bosons. This limit
is approached as interactions relative to kinetic energy become
small. In the limit of strong interactions, the Gutzwiller ansatz can predict
different insulating phases. Depending on the problem, the insulating states
predicted by the Gutzwiller ansatz can be degenerate and these degeneracies can in principle be broken due to the weak inter-site correlations not encountered for in this approach. This was demonstrated for the square and cubic lattices by treating the kinetic energy terms as perturbations~\cite{Collin2010a}. 

Calculating the energy expectation value $\la\psi| H_T|\psi\ra$, using 
the ansatz in Eq.~(\ref{eq:Guztwiller}), gives us an energy functional in
terms of the unknown (complex) amplitudes $a_{{\bf n}^\mathcal{A}}^{({\bf i})}$
and $b_{n_s^\mathcal{B}}^{({\bf j})}$. This energy functional must then be minimized
to find the ground state. Even though this functional is very complex
and the minimization is not always easy, we have found that standard 
conjugate gradient methods work with few caveats. First, the energy functional
can have many local minima into which the minimization algorithm
can become stuck and consequently fail to converge into the
global minimum. In order to build up confidence
in the results it is important to try different initial
states. Second, the minimization algorithm might have trouble
in converging to the correct phase ordering. For example, complex
amplitudes give rise to different phase factors in the condensate
order parameters and in an energy minima these phase factors should
be properly ordered throughout the lattice~\cite{Wirth2011a}.
If the conjugate gradient method is used as a black box, it might
not converge to optimal phase ordering. To get around this, it is important
to impose different orderings into the initial state of the
minimization routines and finally pick the solution that
has the lowest energy. In the absence of a trap we find the solution in a $4\times 4$ lattice where each sublattice has $8$-sites. We use periodic boundary conditions 
and choose to truncate the Fock state expansion of the Gutzwiller ansatz so that the maximum onsite occupation number is $8$.

In the superfluid region we find that the ground state 
phases of the condensate order parameters are arranged 
in the same way as discussed by Wirth {\it et al.}~\cite{Wirth2011a}
for an isotropic lattice. Here, the phase of the condensate order parameter
in the $\mathcal{B}$-sites changes by $\pm 2\pi$ as one moves
around $\mathcal{B}$-sublattice plaquettes. Neighboring plaquettes have an opposite
phase winding. In the $\mathcal{A}$-sites, the onsite order parameters are superpositions of $x$- and $y$-orbitals. These superposition are vortex-like states
proportional to $e^{i\phi}\left(x\pm iy\right)$ and the vorticity
has an opposite sign in neighboring $\mathcal{A}$-sites so that onsite
angular momenta are ordered ``anti-ferromagnetically''. Similarly to
the $\mathcal{B}$-sites, the phase of the 
phase factor $e^{i\phi}$ varies by $\pm 2\pi$ as one travels around $\mathcal{A}$-site plaquettes and neighboring plaquettes have an opposite winding of this phase factor. Far in the superfluid phase where the onsite states can be approximated by coherent states, atoms in the $\mathcal{A}$-sites can be pictured as clockwise or anti-clockwise rotating condensates with a quantization $\langle\hat{L}_z\rangle=\pm1$ where $\hat{L}_z$ is the angular momentum operator in the transverse $z$-direction. 

Note that the swapping symmetry of $x$- and $y$-flavor atoms implies a flip of the vorticity in each site. Closer to the insulating phases where interaction begins to dominate, the picture is more complex and the onsite $x$- and $y$-flavor atoms can become highly entangled. While the Gutzwiller ansatz (\ref{eq:Guztwiller}) is not able to predict inter-site entanglement it indeed captures such intra-site entanglement. As an example, looking at the Mott insulating phase with $n^{\mathcal{A}}=3$ 
atoms in the $\mathcal{A}$-sites, the Gutzwiller method gives a degenerate ground states 
in the $\mathcal{A}$-sites. For example, the states with 
$|\psi\rangle_L=\prod_{\bf i\in \mathcal{A}}a_{3,0}^{(\mathbf{i})}|3,0\rangle_{\mathbf{i}}+a_{1,2}^{(\mathbf{i})}|1,2\rangle_\mathbf{i}$ or 
$|\psi\rangle_R=\prod_{\bf i\in \mathcal{A}}a_{0,3}^{(\mathbf{i})}|0,3\rangle_{\mathbf{i}}+a_{2,1}^{(\mathbf{i})}|2,1\rangle_\mathbf{i}$, with $a_{3,0}^{(\mathbf{i})}=a_{0,3}^{(\mathbf{i})}\approx0.6$ and $a_{1,2}^{(\mathbf{i})}=a_{2,1}^{(\mathbf{i})}\approx-0.8$
are degenerate. 
As discussed above, in the Gutzwiller method these two states are 
decoupled in the insulating phase and breaking the degeneracies requires
improved ansatz and/or higher order perturbation theory in tunneling.
It is clear that these two examples of
insulating states are not eigenstates of $\hat{L}_z$.

\begin{figure*}
\includegraphics[width=0.4\textwidth]{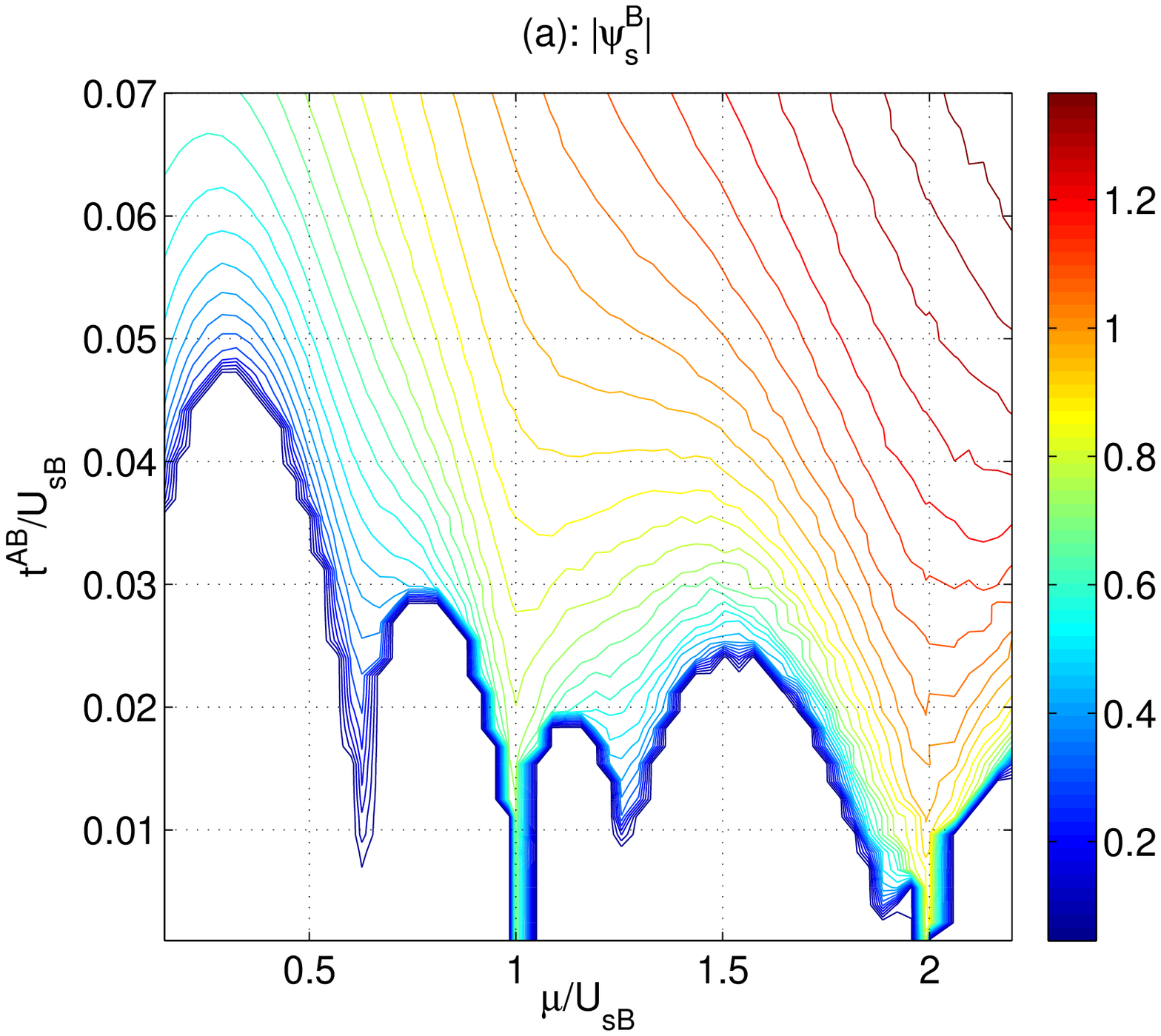}
\includegraphics[width=0.4\textwidth]{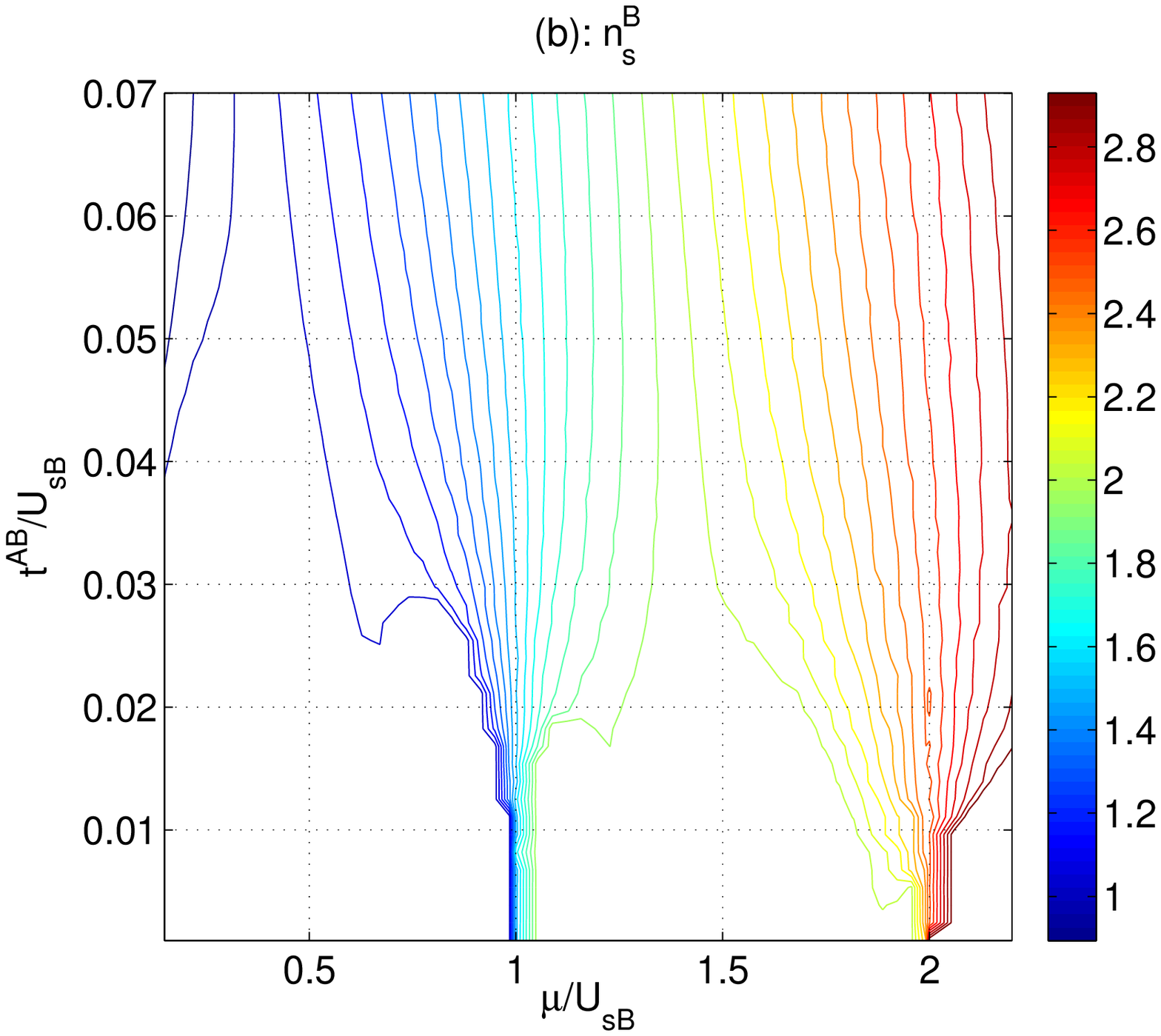}\\
\includegraphics[width=0.4\textwidth]{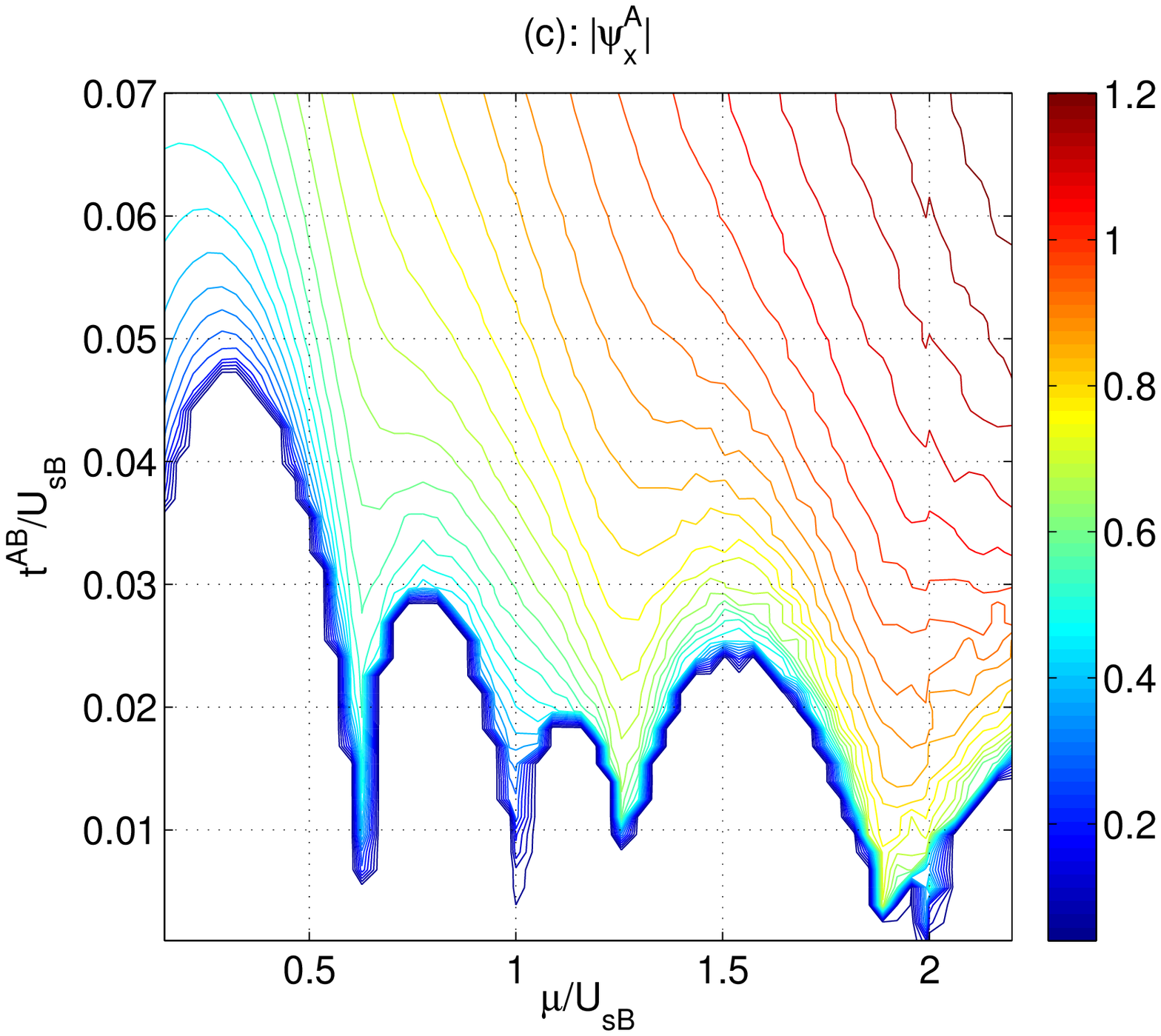}
\includegraphics[width=0.4\textwidth]{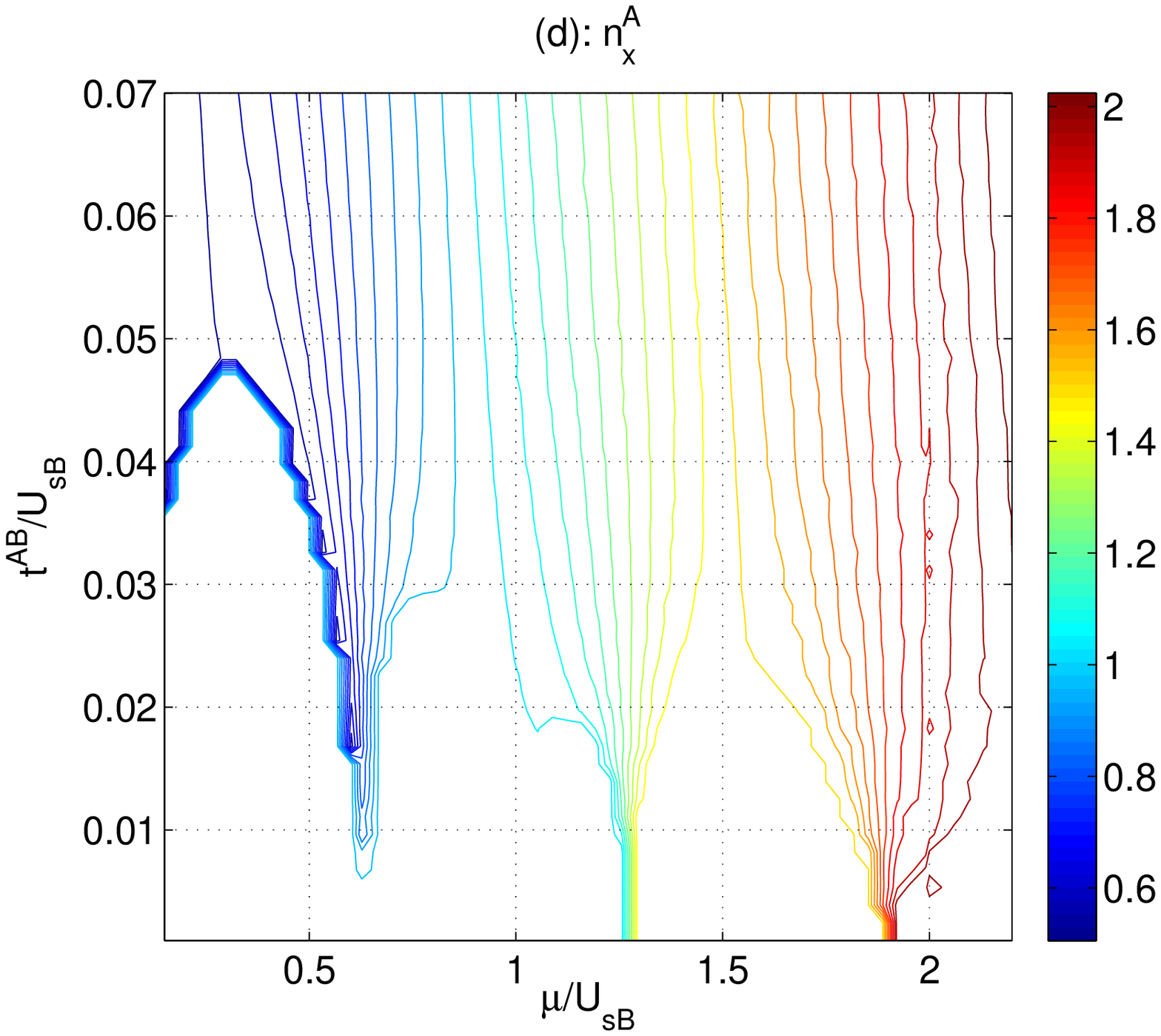}\\
\includegraphics[width=0.4\textwidth]{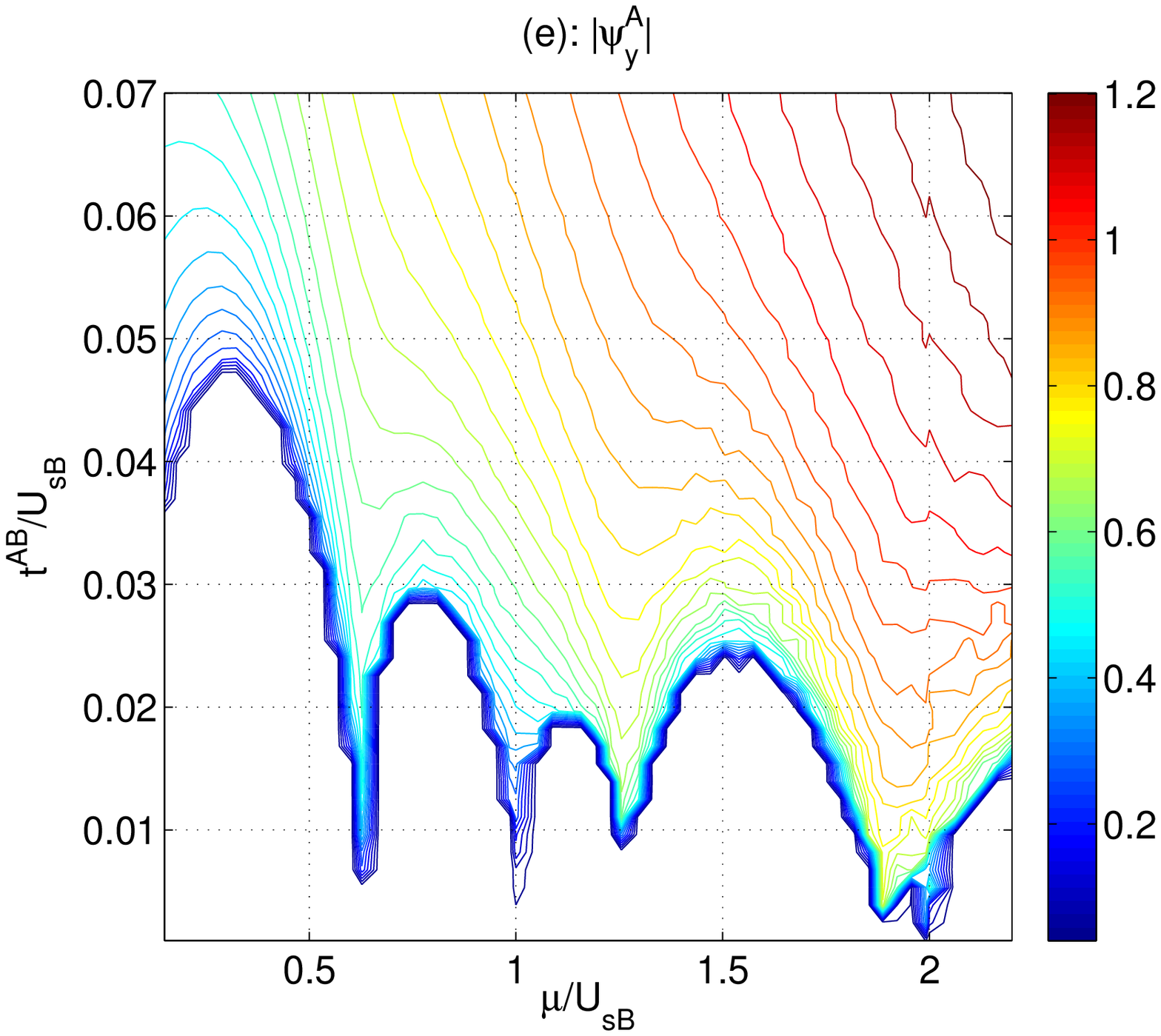}
\includegraphics[width=0.4\textwidth]{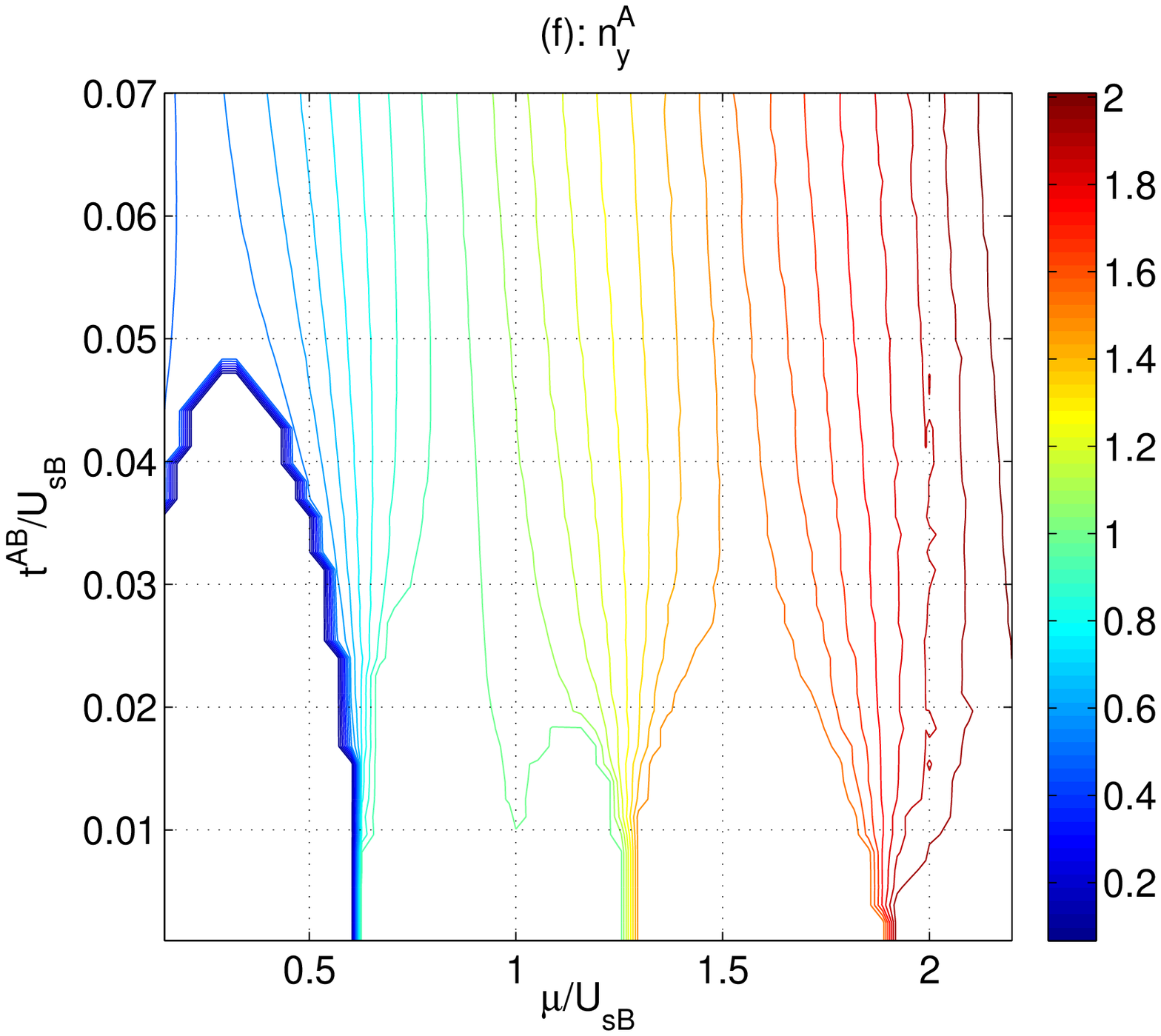}\\
\caption{(Color online) Condensate order parameters and onsite atom numbers
parametrized
by the chemical potential and hybridization tunneling  
$t^{\mathcal{AB}}=t_{xx}^{\mathcal{AB}}=t_{yy}^{\mathcal{AB}}$
when $p$-orbitals are degenerate. (However, in order
to make the plot clearer we did add a very small anisotropy of  $\delta=10^{-4}$
to break the degeneracy of states in $A$-sites with only one
atom per site.) The left hand plots (a),(c), and (e) display condensate densities $|\la\hat\psi_{s,{\bf i}}^\mathcal{B}\ra|^2$ and $|\la\hat\psi_{\beta,{\bf i}}^\mathcal{A}\ra|^2$ ($\beta\in\{x,y\}$), while (b), (d), and (f) show
atom flavor densities $n_{s,{\bf i}}^\mathcal{B}$ and $n_{\beta,{\bf i}}^\mathcal{A}$.  The roughness that is visible especially for higher chemical potentials indicates
the level of numerical uncertainties in these regions. (In the Mott insulating region with 
$n^\mathcal{A}=1$ we choose $n_{x,{\bf i}}^\mathcal{A}=1$, but since interactions
do not contribute here other choices are also possible.)}
\label{fig:isotropic_pd}
\end{figure*}

We show an example of the magnitudes of the
relevant observables in the phase diagram for the isotropic case
with degenerate $p$-orbitals in Fig.~\ref{fig:isotropic_pd}.
As is clear, the phase diagram is very different from the usual
sequence of ever lower Mott-lobes corresponding to higher
onsite atom numbers~\cite{Fisher1989a}. 
In our case there are insulating states with integer
occupation numbers, but since interactions in different sublattices are
different and the other sublattice has several flavors the positions
of the boundaries for different Mott-states are not expected
to be in same positions for different sublattices in the limit
of weak tunneling. The hybridization of orbitals in different
sublattices complicates the picture further.

This interplay between sublattices gives rise to superfluid ``fingers''
extending into the region where each sublattice alone would be
expected to be in a Mott insulator. For example, $\mathcal{B}$-sites make 
a transition from $1$ atom per site to $2$ atoms per site
at $\mu/U_{s\mathcal{B}}=1$. This is apparent in the order parameter
$\la\hat\psi_{s,{\bf i}}^\mathcal{B}\ra^2$ being non-zero in the narrow
region around $\mu/U_{s\mathcal{B}}=1$ even when tunneling becomes weak. With these
parameters and weak tunneling the $\mathcal{A}$-sites are expected to be
in an insulating state with $2$ atoms per site
($|n_x^\mathcal{A}=1,n_y^\mathcal{A}=1\ra$), but coupling with
the condensate order parameter in the $\mathcal{B}$-sites can induce a non-zero
order parameters $\la\hat\psi_{\beta,{\bf i}}^\mathcal{A}\ra$. Similar observations
apply around $\mu/U_{s\mathcal{B}}\approx 1.25$ where the $\mathcal{A}$-sites undergo a transition to $3$ atoms per site. This transition can induce a non-zero condensate order parameter in the $\mathcal{B}$-sites. 

It should be noted that the number fluctuations 
in $x$- and $y$-flavors in the $\mathcal{A}$-sites can be non-zero even
in Mott insulating regions. For example, the Mott insulating
state with $3$ atoms in the $\mathcal{A}$-sites is a superposition
of different basis states with the total of $3$ atom per site. Only the total
number of atoms is fixed to an integer value. The local order parameter breaks
the time-reversal symmetry and the 
angular momentum  in $\mathcal{A}$-sites is non-zero and equal to $\pm 1$
in the condensed region. 
The angular momentum in neighboring $\mathcal{A}$-sites
point in opposite directions. The non-zero value of angular momentum in the 
condensed phase is not surprising since the interaction energy is 
minimized for onsite states with $x\pm i y$ type vortex superpositions
of $p$-orbitals~\cite{Collin2010a,Pinheiro2012a}.

\begin{figure*}
\includegraphics[width=0.4\textwidth]{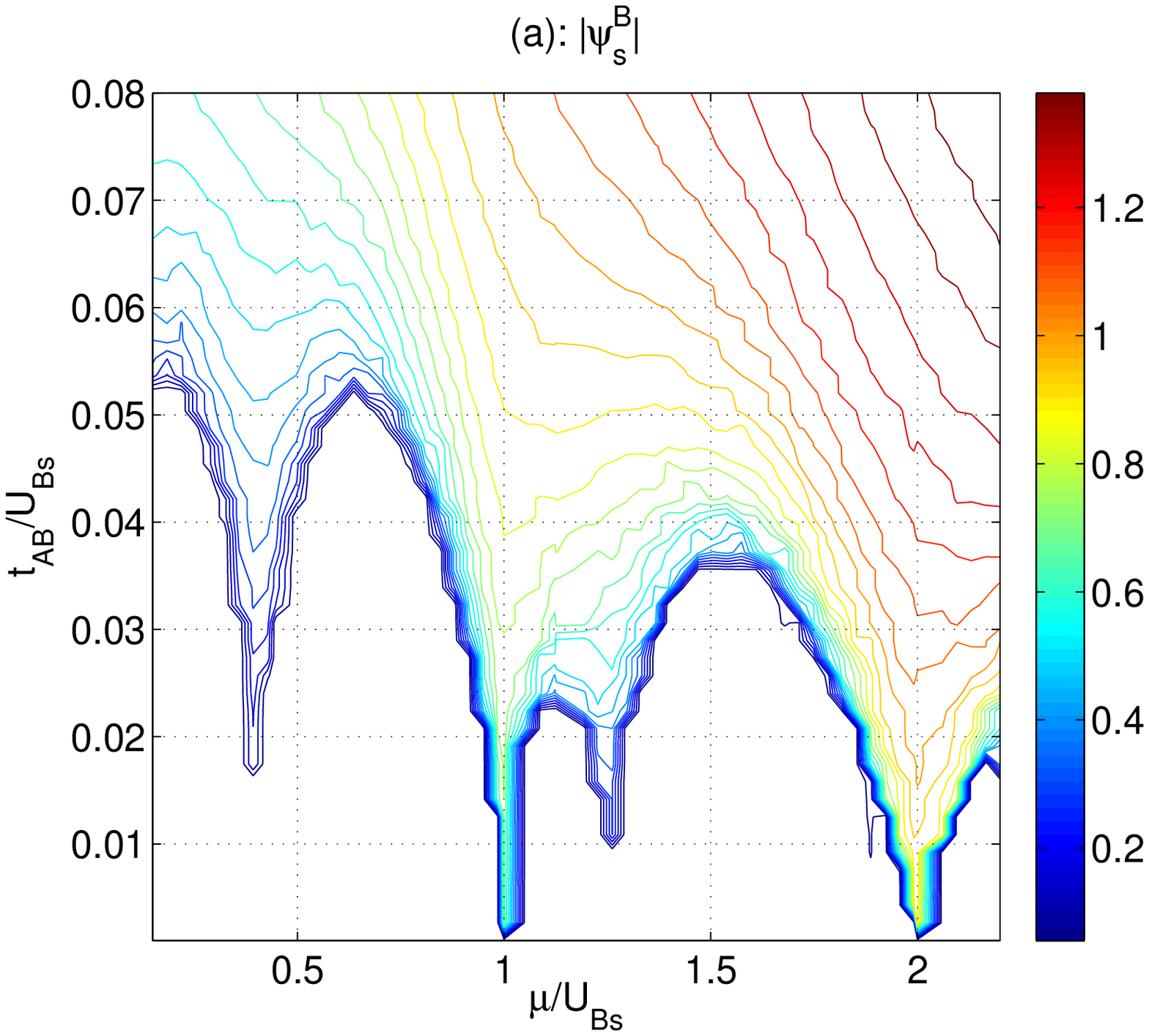}
\includegraphics[width=0.4\textwidth]{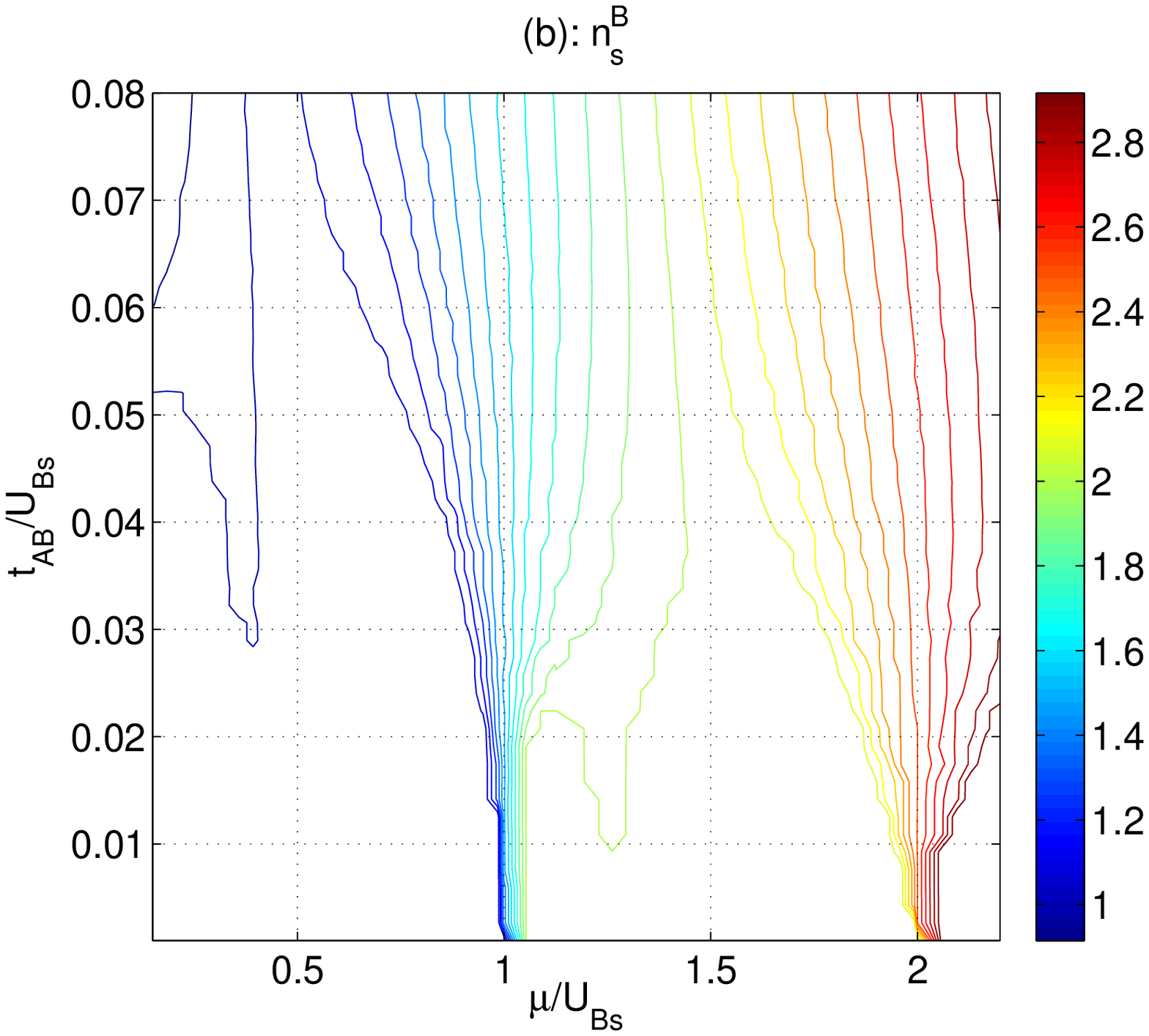}\\
\includegraphics[width=0.4\textwidth]{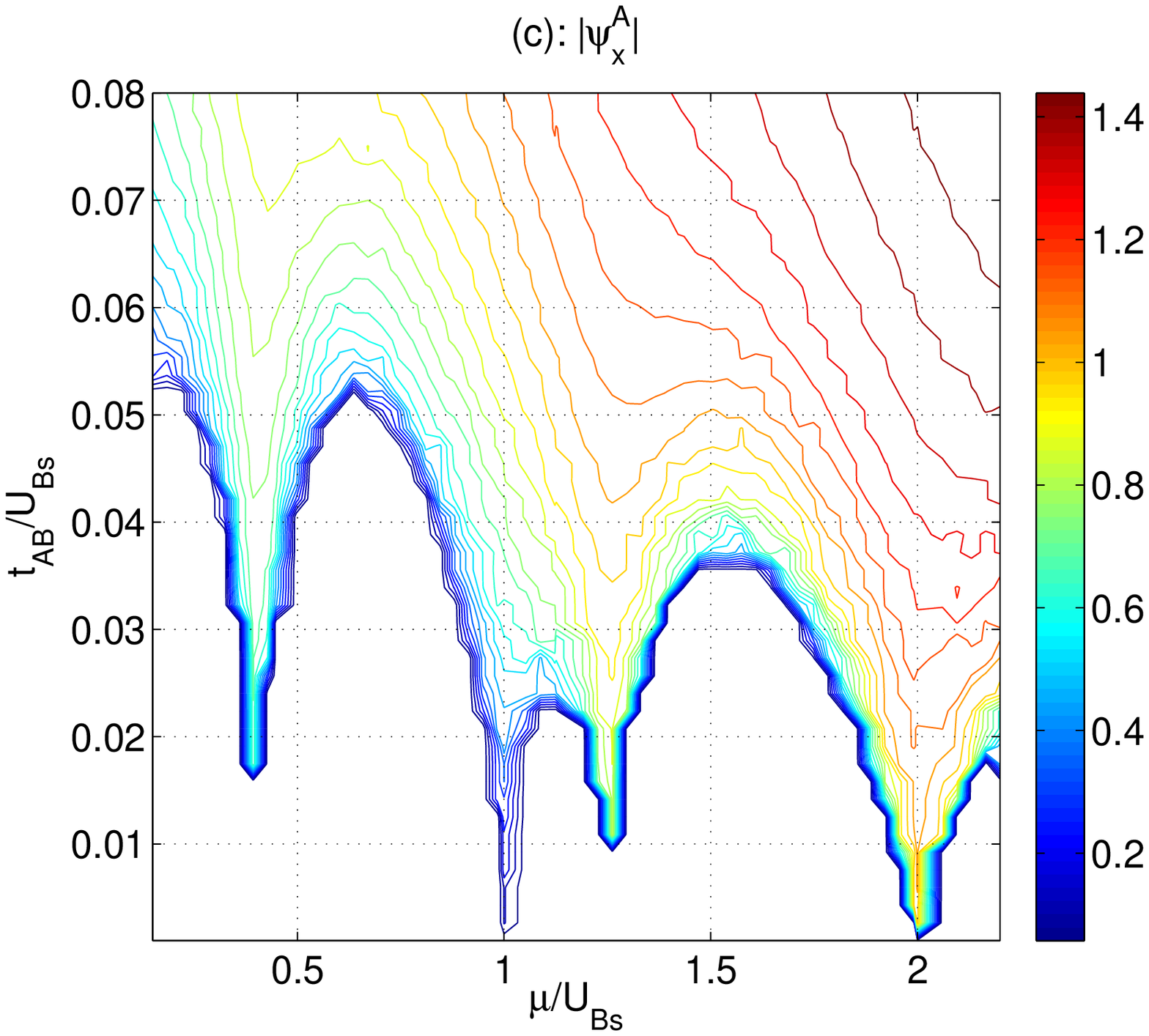}
\includegraphics[width=0.4\textwidth]{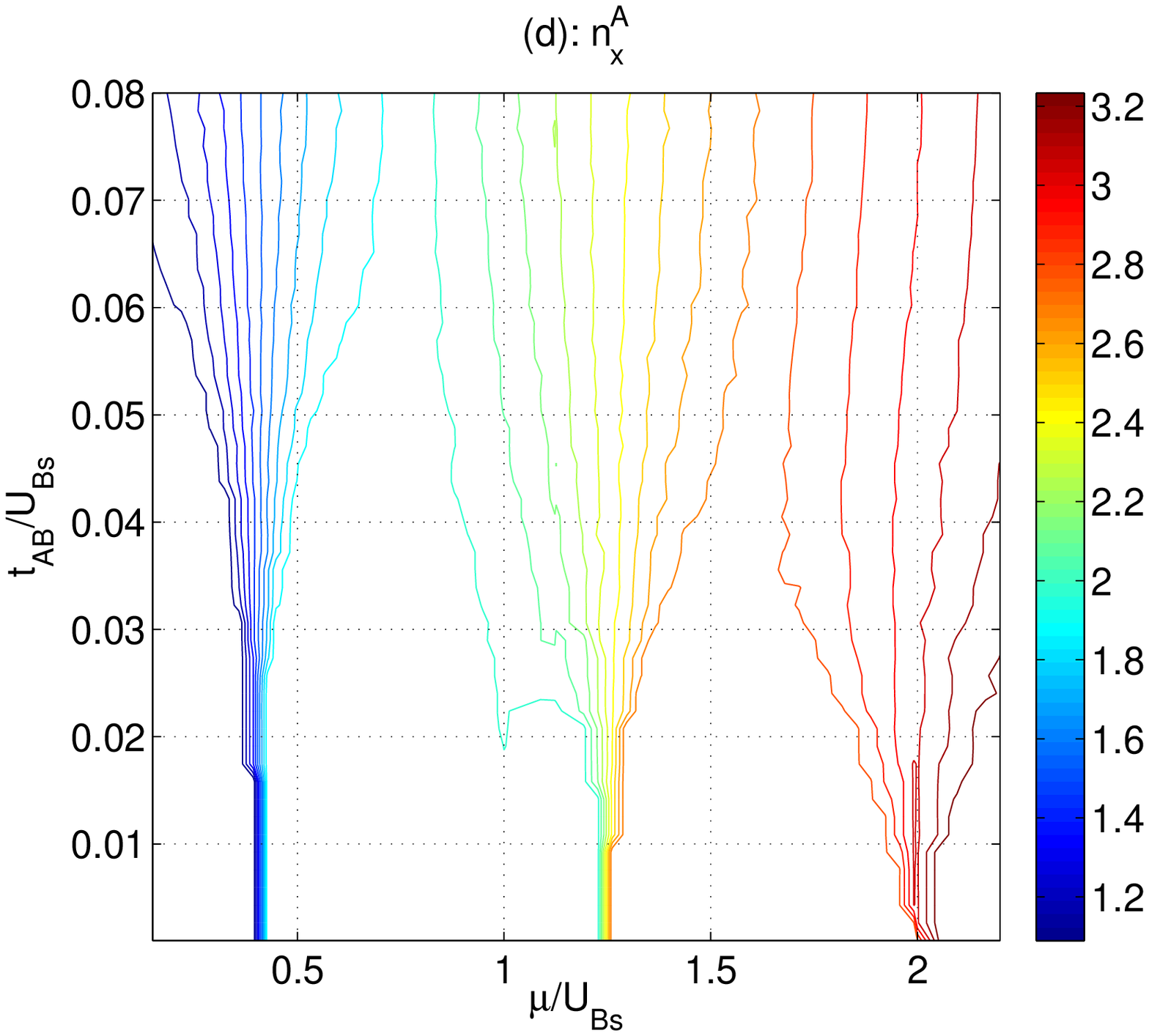}\\
\includegraphics[width=0.4\textwidth]{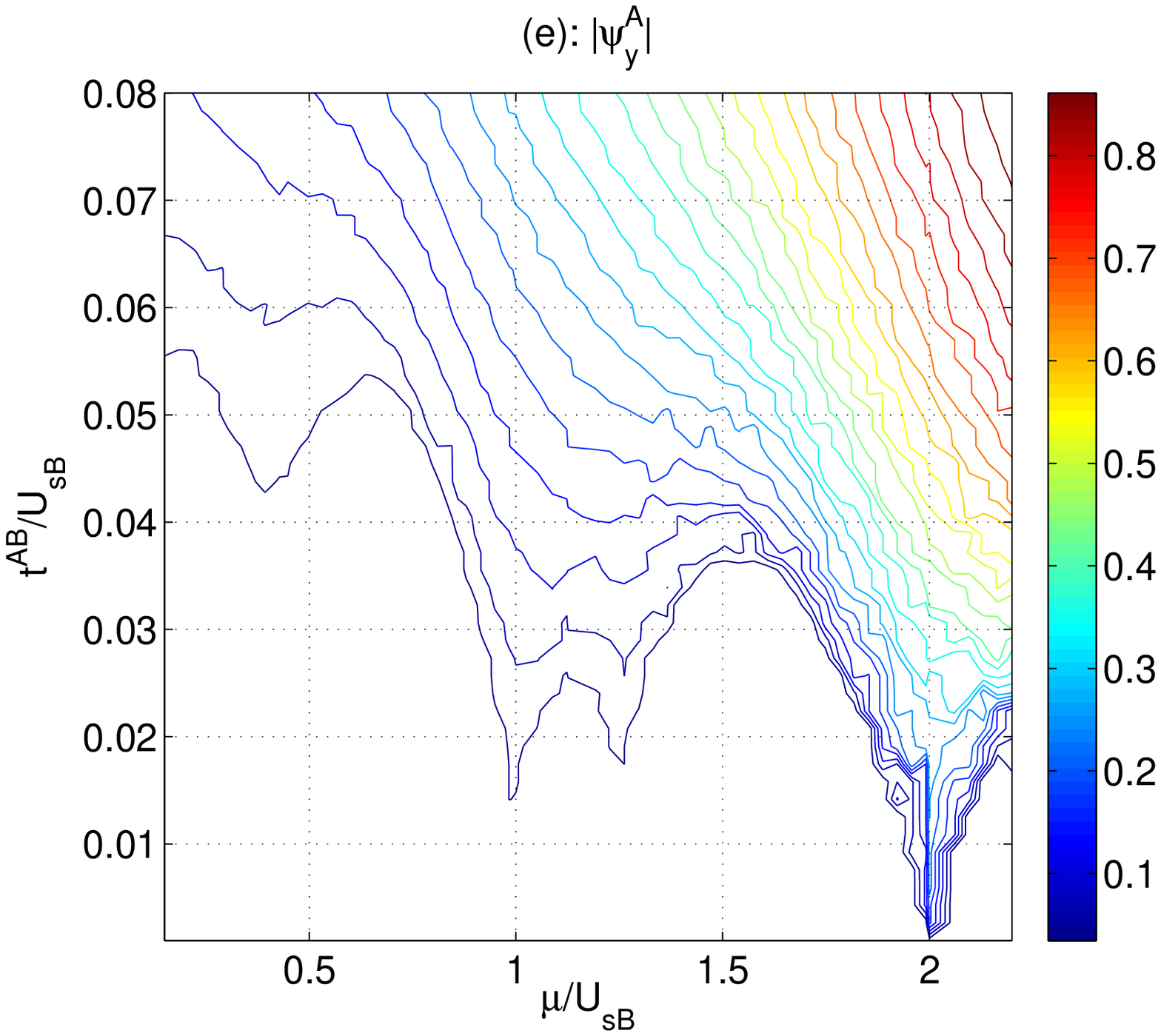}
\includegraphics[width=0.4\textwidth]{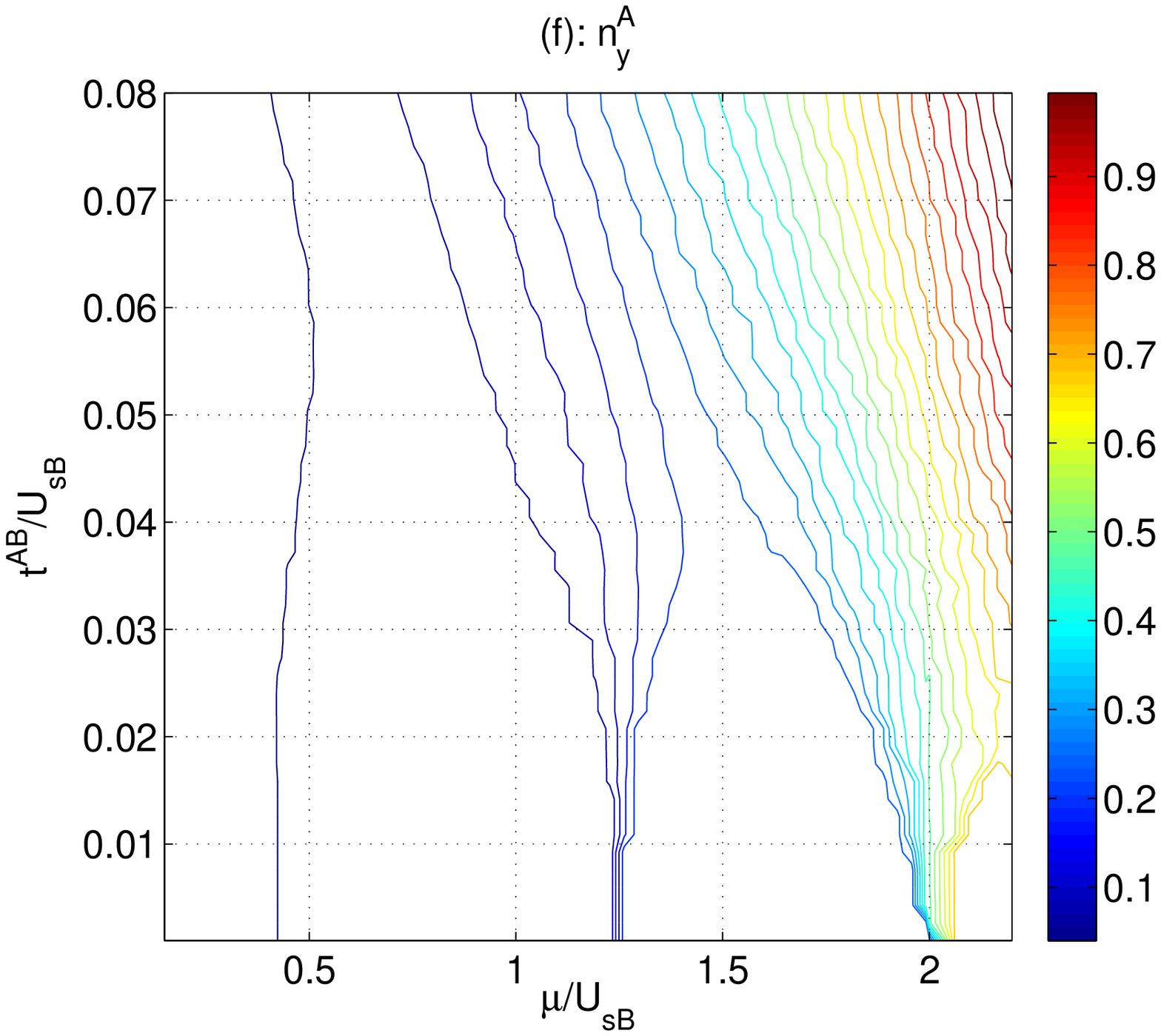}\\
\caption{(Color online) Condensate order parameters and onsite atom numbers
parametrized by the chemical potential and hybridization tunneling  
$t^{\mathcal{AB}}=t_{xx}^{\mathcal{AB}}=t_{yy}^{\mathcal{AB}}$
for the anisotropic case with $\delta/U_{s\mathcal{B}}=1$. 
The plots to the left, (a), (c), and (e), show condensate densities $|\la\hat\psi_{s,{\bf i}}^\mathcal{B}\ra|^2$ and $|\la\hat\psi_{\beta,{\bf i}}^\mathcal{A}\ra|^2$ ($\beta\in\{x,y\}$) while the ones to the right, (b), (d), and (f), display atom flavor densities $n_{s,{\bf i}}^\mathcal{B}$ and $n_{\beta,{\bf i}}^\mathcal{A}$. Small amount of scatter visible especially in (e), is indicative of numerical
uncertainties.}
\label{fig:anisotropic_pd}
\end{figure*}

In Fig.~\ref{fig:anisotropic_pd} we show an example of the phase diagram
for the anisotropic case with a $p$-orbital splitting
$\delta/U_{sB}=1$. In the superfluid regions both $p$-orbitals 
are non-zero, but the order parameter (and density)
for the $y$-orbital is smaller in magnitude. In the Mott insulating 
regions with $1$ or $2$ atoms per site, the onsite interactions (with
these parameters) are
not strong enough to induce large fraction of atoms into the higher $y$-orbitals and
therefore only the $x$-orbitals are substantially populated. However, as the atom number
in the $\mathcal{A}$-sites increases to $3$ or more also the $y$-orbital population becomes 
substantial. 

\begin{figure}
\includegraphics[width=0.5\textwidth]{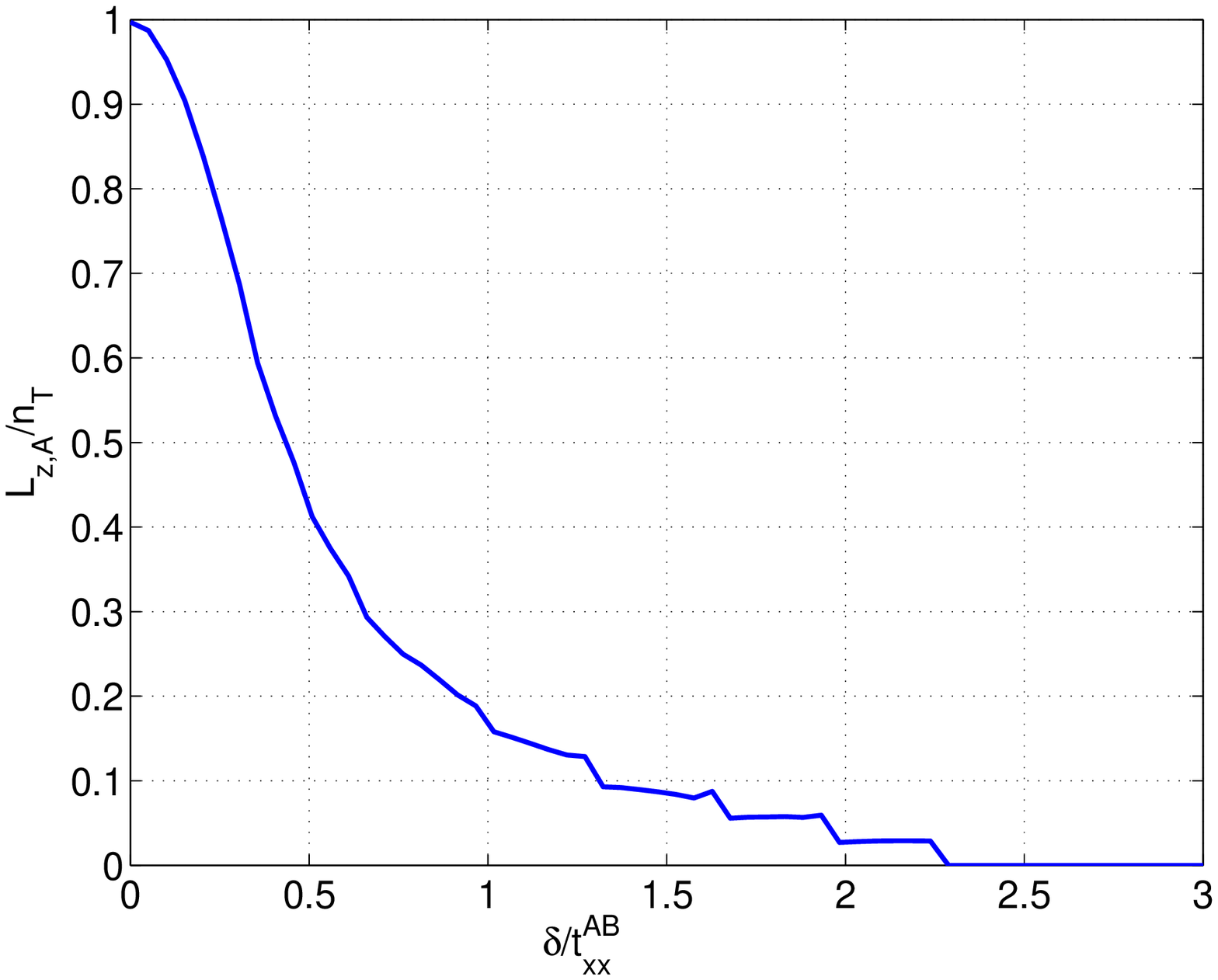}
\caption{Angular momentum per particle in the $\mathcal{A}$-sites
as a function of energy difference $\delta$ between $p$-orbitals.
We choose $t_{xx}^{\mathcal{AB}}/U_{s\mathcal{B}}=0.2$, $\mu/t_{xx}^{\mathcal{AB}}=1$, and $t_{xx}^{\mathcal{AB}}$ as the unit of energy. (The staircase structure at larger $\delta$ is due to numerical limitations in finding the global energy minimum for larger onsite atom numbers with a finite basis set.)
}
\label{fig:deltasweep}
\end{figure}

When we choose $\delta \neq 0$ we break the degeneracy of the $x$- and $y$-orbitals.
In the limit of zero tunneling we expect that if splitting becomes
in some sense large relative to onsite interactions, atoms would prefer
to reside on the $x$-orbital only. It is easy to show that with $2$ atoms per $\mathcal{A}$-site, the transition occurs at $\delta=U_{xx}/3$. It is important to keep in mind that for the case of non-zero tunneling, the situation becomes much more complex and the results may actually depend on the system size. With the Gutzwiller ansatz we find that in the superfluid regime  
(we typically had $t_{xx}^{\mathcal{AB}}/U_{s\mathcal{B}}\sim 0.2\ldots 0.5$), 
the onsite angular momentum (which vanishes if only one orbital is occupied) per particle
is smoothly reduced from its value $\pm 1$ at $\delta=0$ to zero. 
This is demonstrated in Fig.~\ref{fig:deltasweep}.
Vanishing onsite angular momentum is reached when $\delta/t_{xx}^{\mathcal{AB}}\sim 2$ which corresponds fairly well to what would be predicted from the onsite results with a particle number fixed to an integer value (remember that $U_{s\mathcal{B}}\approx U_{xx}$) 
\beq 
\delta/t_{xx}^{\mathcal{AB}}=
\left(\delta/U_{s\mathcal{B}}\right)\times \left(U_{s\mathcal{B}}/t_{xx}^{\mathcal {AB}}\right)
\approx \frac{1}{3}\times \left(U_{s\mathcal{B}}/t_{xx}^{\mathcal{AB}}\right).
\enq
As expected, the onsite angular momentum also drops faster for larger $t_{xx}^{\mathcal{AB}}/U_{s\mathcal{B}}$ since this implies smaller onsite interaction strengths.

If we replace the operators with complex numbers $\psi_\alpha$ to derive a
Gross-Pitaevskii equations for each orbital, we find that for the onsite 
problem the effective 
chemical potential and thus also the density of $y$-orbitals vanish
when $\delta/2=\mu-U_{xy}n_x$ at which point the density of the
$x$-orbital is related to the chemical potential through
$n_x=(\mu+\delta/2)/U_{xx}$. This implies that in this limit the transition
from states with orbital angular momentum to pure $x$-orbital 
condensate happens at $\delta_c=(U_{xx}-U_{xy})n_x$, where $n_x=|\psi_x|^2$.

\subsection{Trapped system}
\label{sec:trappedsystem}
Typical experiments would most likely involve the presence of a confining trapping potential and for this reason
it is important to also discuss the behavior with inhomogeneous density
distributions. Our predictions for the phase diagram in a homogeneous 
system suggest an interesting possibility in a trap. Usually the solution
of the Bose-Hubbard model in a trap gives rise ``a wedding cake'' structure
where Mott plateaus corresponding to different 
integer fillings are sandwiched between superfluid regions~\cite{batrouni_mott_2002}.

If we were to apply a local density approximation to our system, we could
think of the chemical potential as a local quantity
$\mu=\mu_{center}-V_{trap}(i_x,i_y)$, 
where $V_{trap}(i_x,i_y)$ would typically be a harmonic trap.
Traversing from the center of the cloud to its edge would correspond
to moving in the phase diagram from some high value of $\mu/U_{s\mathcal{B}}$
towards zero. If the starting point is in the Mott insulating phase
we could indeed have a wedding cake structure for each sublattice, but
their Mott plateaus do not always coincide. Furthermore, we can have
situations when a condensate order parameter appears inside
the same Mott plateau. We will next demonstrate that
these simple observations are valid in a trap also beyond the local density approximation.

We can do this within the theoretical framework used so far, but
replacing the chemical potential $\mu$ with $\mu_{center}-V_{trap}(i_x,i_y)$
in the Hamiltonian in Eq.~(\ref{eq:H0}) and then solving the problem
with the trapping potential 
\beq
V_{trap}(i_x,i_y)\!=\!\gamma\!\left[\!(i_x\!-\!(N_x\!+\!1)/2)^2+(i_y\!-\!(N_y\!+\!1)/2)^2\right]
\enq
with $N_x$ and $N_y$ being the number of sites along $x$ and $y$ respectively. 
(The minima of the harmonic potential is shifted to $((N_x+1)/2,(N_y+1)/2)$
since we choose $i_\alpha\in \{1\ldots N_\alpha\}$.) 
As an example, we choose an isotropic lattice with $t/U_{s\mathcal{B}}=0.015$ and 
the chemical potential in the center $\mu_{center}/U_{s\mathcal{B}}=1.5$ so that
in the center of the cloud we expect the $\mathcal{A}$-sites to be in an insulating state with three atoms per site. The trap coefficient $\gamma$ we choose
in such a way that $\mu_{center}-V_{trap}(i_x,i_y)$ becomes
negative at the edge of the lattice so that the density vanishes there. 

\begin{figure*}
\includegraphics[width=0.4\textwidth]{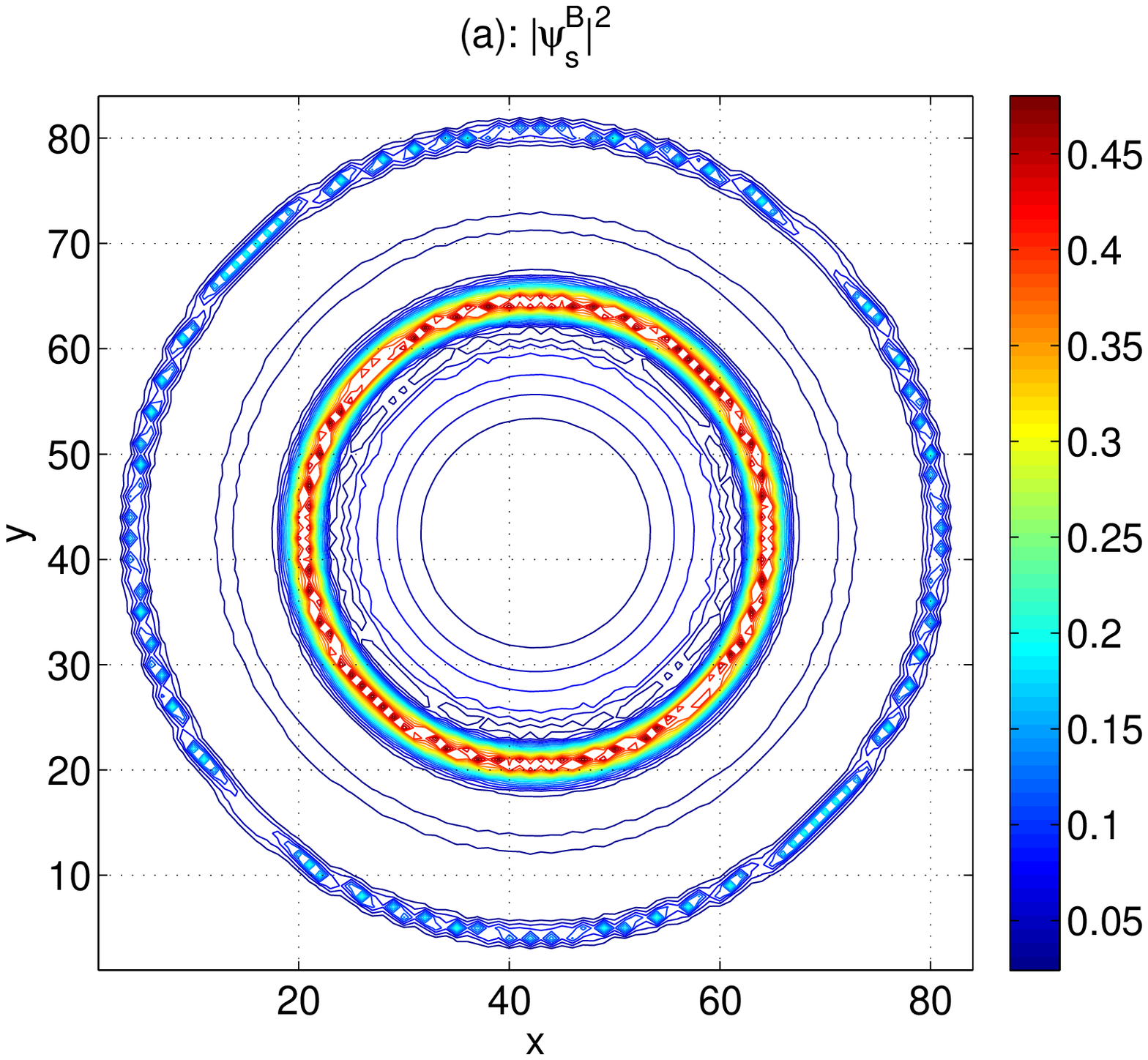}
\includegraphics[width=0.4\textwidth]{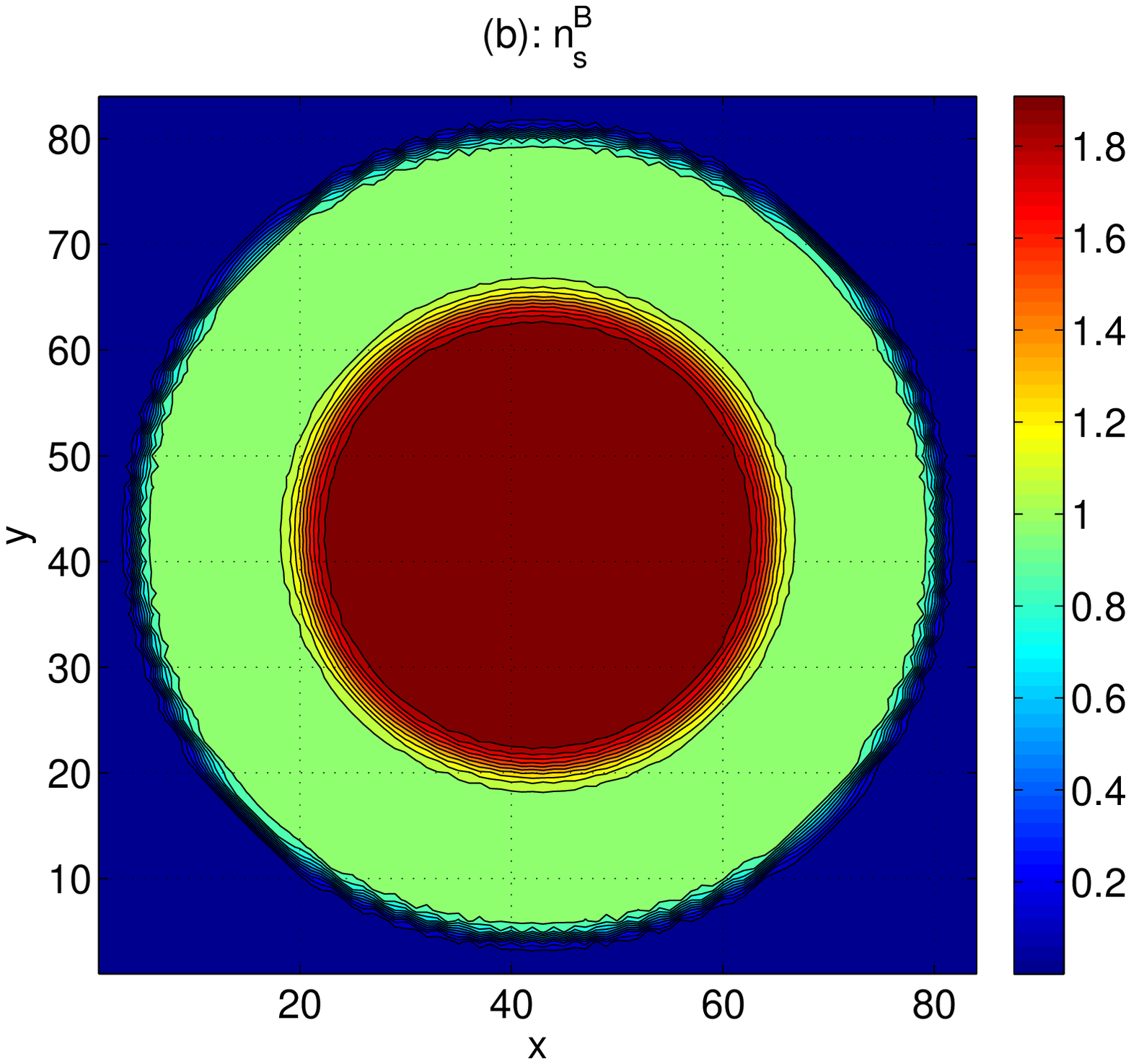}\\
\includegraphics[width=0.4\textwidth]{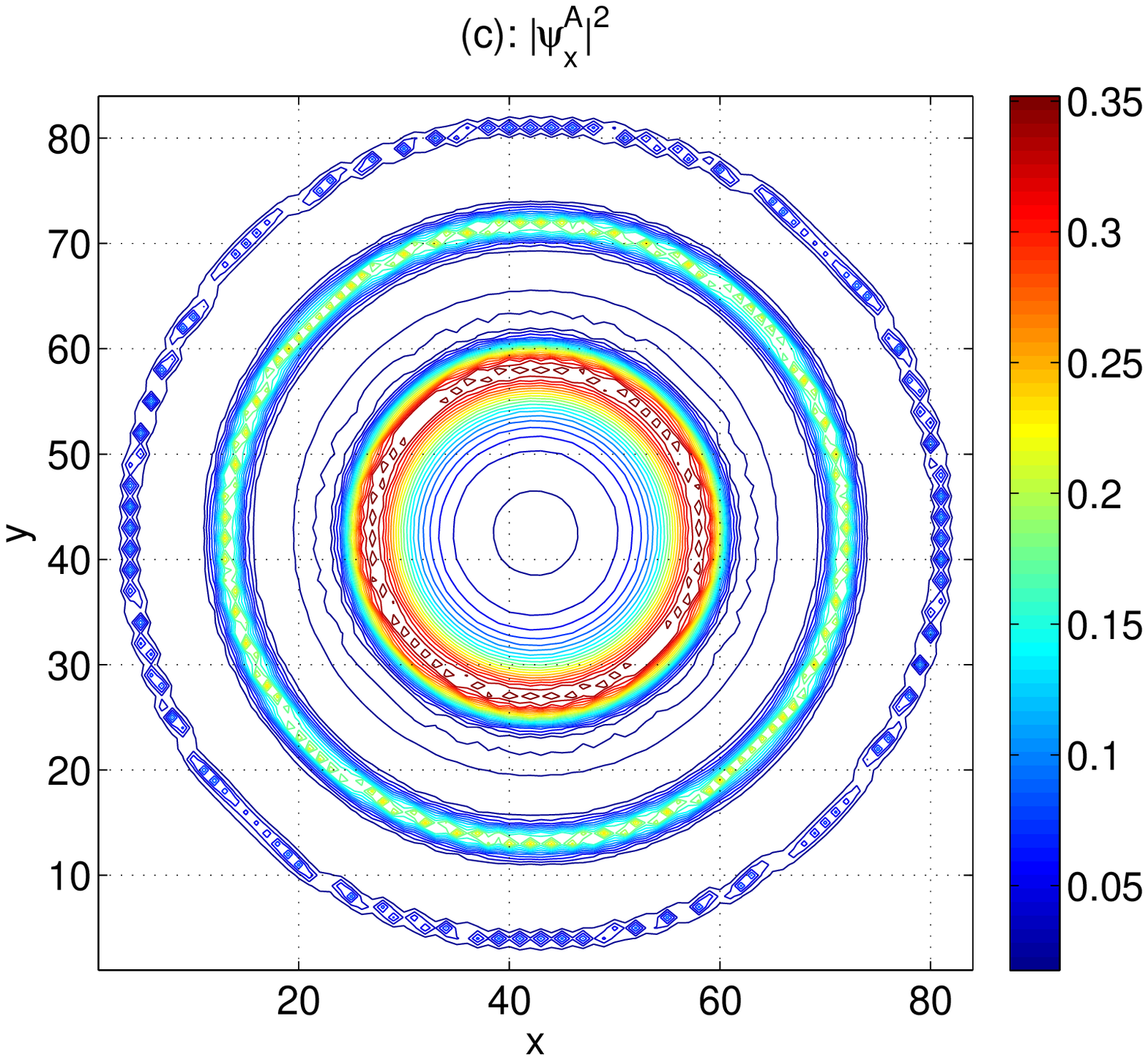}
\includegraphics[width=0.4\textwidth]{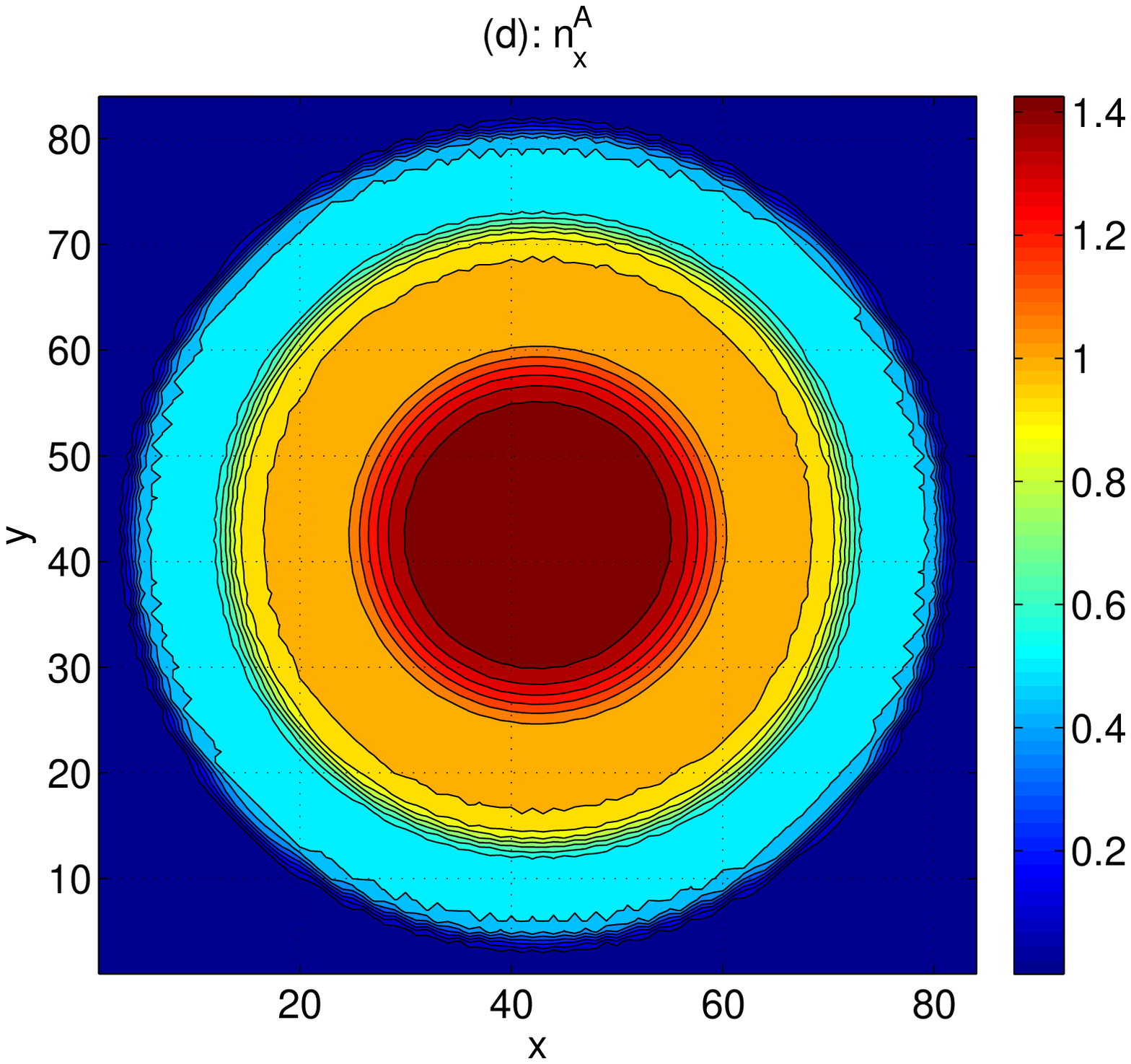}
\caption{(Color online) Condensate and flavor densities in a trap. The left hand plots (a) and (b) give the condensate densities $|\la\hat\psi_{s,{\bf i}}^\mathcal{B}\ra|^2$
and $|\la\hat\psi_{x,{\bf i}}^\mathcal{A}\ra|^2$ while (c) and (d) show
atom flavor densities $n_{s,{\bf i}}^\mathcal{B}$ and $n_{x,{\bf i}}^\mathcal{A}$. We choose $t/U_{s\mathcal{B}}=0.015$,  $\mu_{center}/U_{s\mathcal{B}}=1.5$, and $\gamma$
in such a way that the density vanishes at the edge of the lattice. Since the 
lattice is isotropic the densities for the $y$-orbital are the
same as for the $x$-orbital and are not plotted here. The axes give the lattice sites in the two laboratory directions. (Plotted quantities are only defined in their respective sublattices. However, to make the figure clearer we filled in the relevant 
values also to the other sublattice by taking the average over the 
$4$ neighboring sites.)
}
\label{fig:trapped}
\end{figure*}

We demonstrate the resulting ground state of the trapped bosons 
in Fig.~\ref{fig:trapped}. The bosons arrange themselves into the
familiar wedding cake structure with Mott-insulating regions separated
by superfluid-regions. Remarkably, as suggested by the 
results in the absence of trapping potential, since our system has two 
different sublattices with different onsite interactions, superfluid
"rings'' can occur in different locations for different orbitals.
For example, closest to the center we have a region where the $\mathcal{A}$-sites
are Mott-insulators with $n^\mathcal{A}=n_x^\mathcal{A}+n_y^\mathcal{A}=3$ while the 
$\mathcal{B}$-sites are insulating with $n^\mathcal{B}=2$. 
The transition to $n^\mathcal{A}=2$ 
phase occurs via a superfluid phase in the $\mathcal{A}$-sites. However, in this region 
the $\mathcal{B}$-sites are still very small. 
Also, there is a condensed phase between regions 
with $n^\mathcal{B}=2$ and $n^\mathcal{B}=1$ while the condensate order parameters
in $\mathcal{A}$-sites are negligible.
Consequently, the physics predicted by using the theory without the trapping potential
can also persist in trapped systems. 

Recently, the trapped system of $p$-band bosons in a square lattice was analyzed and it was found that the density of different $x$- or $y$-orbital atoms were elongated in one direction and the symmetry of the confining trap was broken~\cite{Pinheiro2012a}. 
The present system is different due to the hybridization of $s$- and $p$-orbitals, 
which implies that the condensate cloud preserves the symmetry for an isotropic trap. 
On the other hand, if one prepares the system so that the tunneling coefficients 
$t_{xx}^{\mathcal{AB}}$ and $t_{yy}^{\mathcal{AB}}$
are unequal in magnitude, similar anisotropies might be expected also here.

\section{Conclusions}
\label{sec:conclusions}
In this paper we have derived a TB model to describe
ultracold atoms in a bipartite optical lattice with three
hybridized orbitals. We have also solved
the resulting generalized Bose-Hubbard model and found strong modifications
to the Mott insulator superfluid phase diagram which is found in the simplest
lowest band Bose-Hubbard model. Novel phenomena was also demonstrated for the confined system that includes a harmonic
trap. From that solution we 
found that the unusual phase diagram of the multi-band Bose-Hubbard model
can be reflected as possessing non-trivial wedding cake structure of Mott insulating
regions for different sublattices. In particular, a non-zero condensate order parameter in 
one sublattice can coexist with a Mott plateau in another sublattice
and also appear inside the same Mott plateau.
Such effects are observable since Mott insulating regions can be detected
{\it in-situ} and atoms in optical lattices can be manipulated even at a single site 
resolution~\cite{Bakr2009a,Gemelke2009a,Sherson2010a,Weitenberg2011a}. Furthermore, since different sublattices have different atom-atom 
interactions the states with more than one atom per site would generally give rise
to different mean-field shifts if transitions to other hyperfine states are considered.
This suggest a possibility of addressing different sublattices with 
microwave fields of different frequencies, for example.

In this paper we have not addressed the dynamical behavior of
bosons in a bipartite lattice. However, using the theoretical framework 
derived here that would be not only doable, but also interesting since 
in the experiments conducted so far bosons have been initially
prepared in an excited state whose dynamical behavior is poorly
understood. 

\begin{acknowledgments}
Financial support from the Swedish Research Council (Vetenskapsr\aa det) is acknowledged. JL acknowledges financial support from DAAD (Deutscher Akademischer Austausch Dienst) and the Royal Research Council Sweden (KVA). JPM acknowledges financial support from the Academy of Finland (Project 135646).
\end{acknowledgments}


\end{document}